\documentclass[10pt,letterpaper]{article}
\usepackage[top=0.85in,footskip=0.75in]{geometry}
\pdfoutput=1
\usepackage{changepage}
\usepackage[utf8]{inputenc}
\usepackage{textcomp,marvosym}
\usepackage{fixltx2e}
\usepackage{amsmath,amssymb}
\usepackage{nameref,hyperref}
\usepackage[right]{lineno}
\usepackage{microtype}
\DisableLigatures[f]{encoding = *, family = * }
\usepackage{rotating}
\usepackage{todonotes}
\usepackage{nameref}

\usepackage[aboveskip=1pt,labelfont=bf,labelsep=period,justification=raggedright,singlelinecheck=off]{caption}
\makeatletter
\renewcommand{\@biblabel}[1]{\quad#1.}
\makeatother
\reversemarginpar

\usepackage{lastpage,fancyhdr,graphicx}
\usepackage{epstopdf}
\DeclareRobustCommand{\greektext}{%
  \fontencoding{LGR}\selectfont\def\encodingdefault{LGR}}
\DeclareRobustCommand{\textgreek}[1]{\leavevmode{\greektext #1}}
\DeclareFontEncoding{LGR}{}{}
\DeclareTextSymbol{\~}{LGR}{126}
\usepackage{subfig}

\begin{document}
\vspace*{0.35in}

\begin{flushleft}
{\Large
\textbf\newline{Dynamic Decomposition of Spatiotemporal Neural Signals}
}
\newline
\\
Luca Ambrogioni\textsuperscript{1},
Marcel A. J. van Gerven\textsuperscript{1},
Eric Maris\textsuperscript{1}
\\
\bigskip
\bf{1} Radboud University, Donders Institute for Brain, Cognition and Behaviour, Nijmegen, The Netherlands 
\\
\bigskip
* l.ambrogioni@donders.ru.nl

\end{flushleft}

\section*{Abstract}
Neural signals are characterized by rich temporal and spatiotemporal dynamics that reflect the organization of cortical networks. Theoretical research has shown how neural networks can operate at different dynamic ranges that correspond to specific types of information processing. Here we present a data analysis framework that uses a linearized model of these dynamic states in order to decompose the measured neural signal into a series of components that capture both rhythmic and non-rhythmic neural activity. The method is based on stochastic differential equations and Gaussian process regression. Through computer simulations and analysis of magnetoencephalographic data, we demonstrate the efficacy of the method in identifying meaningful modulations of oscillatory signals corrupted by structured temporal and spatiotemporal noise. These results suggest that the method is particularly suitable for the analysis and interpretation of complex temporal and spatiotemporal neural signals.

\section*{Introduction}
Human neocortex has an impressively complex organization. Cortical
electrical activity is determined by dynamic properties of neurons
that are wired together in large cortical networks. These neuronal
networks generate measurable time series with characteristic spatial and
temporal structure. In spite of the staggering complexity of cortical
networks, electrophysiological measurements can often be properly
described in terms of a few relatively simple dynamic components.
By dynamic components we mean signals that exhibit characteristic properties
such as rhythmicity, time scale and peak frequency. For example, neural oscillations
at different frequencies are extremely prominent in electroencephalographic
(EEG) and magnetoencephalographic (MEG) measurements and have been
related to a wide range of cognitive and behavioral states \cite{cheyne2013meg,roux2014working,basar2013review}.
Neural oscillations have been the subject of theoretical and experimental
research as they are seen as a way to connect the dynamic properties
of the cortex to human cognition \cite{tallon1999oscillatory,engel2010beta,kirschfeld2005alpha,klimesch2007eeg,jensen2010shaping}. Importantly,
an oscillatory process can be described using simple mathematical
models in the form of linearized differential equations \cite{bressloff2011spatiotemporal}. 

In this paper, we introduce a framework to integrate prior knowledge
of neural signals (both rhythmic and broadband) into an analysis framework based on Gaussian process (GP) regression \cite{rasmussen2006gaussian}. The aim is to decompose the measured time series
into a set of dynamic components, each defined by a linear stochastic
differential equation (SDE). These SDEs determine a prior probability
distribution through their associated GP covariance functions. The
covariance function specifies the prior correlation structure of the
dynamic components, i.e. the correlations between the components'
activity at different time points. Using this prior, a mathematical
model of the signal dynamics is incorporated into a Bayesian data
analysis procedure. The resulting decomposition method is able to
separate linearly mixed dynamic components from a noise-corrupted
measured time series. This is conceptually different from blind decomposition
methods such as ICA and PCA \cite{comon1994independent,tipping1999probabilistic} that necessarily rely on the
statistical relations between sensors and are not informed by a prior
model of the underlying signals. In particular, since each component extracted using the GP decomposition is obtained from an explicit model of the underlying process, these components are easily interpretable and can be naturally compared across different participants and experimental conditions.

The GP decomposition
can be applied to spatiotemporal brain data by imposing a spatial
smoothness constraint at the level of the cortical surface. We will show that the resulting
spatiotemporal decomposition is related to well-known source
reconstruction methods \cite{hamalainen1994interpreting,pascual2002functional,tarantola2005inverse,petrov2012harmony} and allows to
localize the dynamic components across the cortex. The connections
between EEG/MEG source reconstruction and GP regression have recently been shown by Solin et al.~\cite{solin2016gprec}. Our approach complements and extends their work by introducing an explicit additive model of the underlying neural dynamics.

Through computer simulations and analysis of empirical data, we show
that the GP decomposition method allows to quantify subtle modulations of
the dynamic components, such as oscillatory amplitude modulations,
and does so more reliably than conventional methods. We also demonstrate that the output of the method is highly interpretable and can be effectively used for uncovering reliable spatiotemporal phenomena in the neural data. Therefore, when applied to the data of a cognitive experiment, this approach may give rise to new insights into how cognitive states arise from neural dynamics.

\section*{Results}
In the following, we will show how to construct a probabilistic model of the neural dynamics that captures the main dynamical features of the electrophysiological signals. The temporal dynamics of the neural sources are modeled using linear SDEs, and these in turn determine a series of GP prior distributions. These priors will be used to decompose the signal into several dynamic components with a characteristic temporal correlation structure. Building from the temporal model, we introduce a spatiotemporal decomposition method that can localize the dynamic components on the cortical surface.

\subsection*{Decomposing a signal using temporal covariance functions}


\paragraph*{Modeling neural activity with stochastic differential equations}

\begin{figure}[]
\centering
    	\includegraphics[] {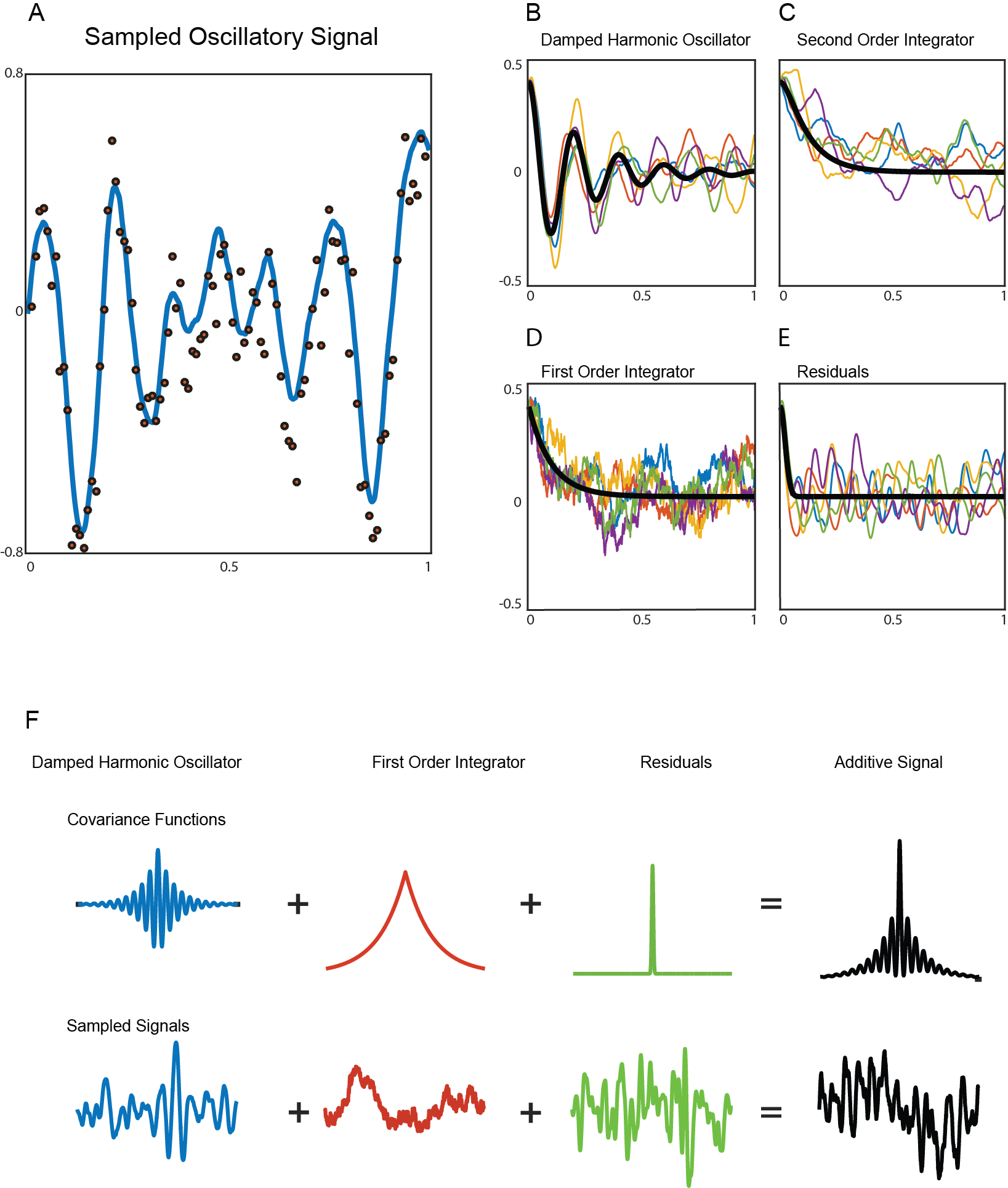}
\caption{\bf Stochastic processes and covariance functions.}
A) Example of a continuous-time oscillatory process (blue line) sampled at discrete equally-spaced time points though noise corrupted measurements (red dots). B--E) Samples (colored) and expected values (black) of the stochastic processes. The processes are a damped harmonic oscillator, second order integrator, first order integrator and residuals respectively. The samples start from an excited state and decay back to their respective stationary distribution. F) Illustration of the decomposition of a complex signal's covariance function into simpler additive components. This corresponds to an additive decomposition of the measured time series.  The second order integrator process has been excluded from this panel for visualization purposes. 
\label{fig1}
\end{figure}

We start our exposition by considering a single sensor that measures
the signal produced by the synchronized subthreshold dynamics of some
homogeneous neuronal population. Neural activity is defined for all possible time points. However,
it is only observed through discretely-sampled and noise-corrupted
measurements $y_t$. We assume the observation noise
$\xi(t)$ to be Gaussian but not necessarily white. The effect
of the discrete and noise-corrupted sampling is exemplified in Fig.~\ref{fig1}A, which shows a simulation of a continuous-time process sampled at regular intervals and corrupted by white noise. Modeling the neural signal as a continuous (rather than a discrete) time series has the advantage of being invariant under changes of sampling frequency
and can also accommodate non-equidistant samples. 

Our prior of the temporal dynamics of the neural activity is specified using linear SDEs. For example, we model the neural oscillatory process $\varphi(t)$ using the following equation:
\begin{equation}
\frac{d^{2}}{dt^{2}}\varphi(t)+b\frac{d}{dt}\varphi(t)=-\omega_{0}^{2}\varphi(t)+w(t)\,.\label{eq:oscillator SDE}
\end{equation}
This differential equation describes a damped harmonic oscillator,
which responds to input by increasing its oscillatory amplitude. The
parameter $b$ regulates the exponential decay of these input-driven excitations. The frequency \textgreek{w} of these excitations is equal to $\sqrt{\omega_{0}^{2}-\frac{1}{2}b^{2}}$.
Clearly, this frequency is only defined for $\omega_{0}^{2}>\frac{1}{2}b^{2}$.
For larger values of $b$, the system ceases to exhibit oscillatory responses
and is said to be overdamped. These dynamical states are referred to as
an oscillator in case $\omega_{0}^{2}>\frac{1}{2}b^{2}$ and an integrator
in case $\omega_{0}^{2}<\frac{1}{2}b^{2}$~\cite{izhikevich2007dynamical}. 

We assume the process to be driven by a random input $w(t)$ (also denoted as \emph{perturbation}). This random function models the combined
effect of the synaptic inputs to the neuronal population that generates the signal. Fig.~\ref{fig1}B shows the expected value (black)
and a series of samples (coloured) of the process, starting from an
excited state $(\varphi(0)=0.4)$ and decaying back to its stationary
dynamics. Note that the expected value converges to zero whereas
the individual samples do not; this is due to the continued effect of the random input.
Also note that the samples gradually become phase inconsistent, with
the decay of phase consistency being determined by the damping parameter
$b$. Thus, the damping parameter also determines the decay of the temporal
correlations. 

In general, we model the measured time series as a mixture of four processes, which
we will now describe. Of these four, one reflects rhythmic brain activity (i.e., an oscillation),
two reflect non-rhythmic brain activity, and one accounts for the
residuals:

\begin{itemize}
\item \textit{Damped harmonic oscillator.} Oscillations are a feature
of many electrophysiological recordings~\cite{gray1989stimulus,lubenov2009hippocampal}, and they are thought to
be generated by synchronized oscillatory dynamics of the membrane
potentials of large populations of pyramidal neurons \cite{silva1991intrinsic}. We
model the neural oscillatory process as a stochastic damped harmonic
oscillator as defined in Eq.~(\ref{eq:oscillator SDE}) with damping
coefficient $b < \sqrt{2\omega_{0}^{2}}$. This linear
differential equation can be obtained by linearizing a model of the
neuronal membrane potential that is characterized by sub-threshold
oscillations~\cite{izhikevich2007dynamical}. 
\item \textit{Second order integrator.} We model the smooth non-oscillatory
component of the measured time series using an equation of the same
form as Eq.~(\ref{eq:oscillator SDE}) but in the overdamped state. We will denote this dynamic component as $\chi(t)$. In the overdamped regime, the equation has smooth,
non-rhythmic solutions (see Fig.~\ref{fig1}C). Equations like these emerge
by linearizing neuronal models around a non-oscillatory fixed point~\cite{izhikevich2007dynamical}.
\item \textit{First order integrator.} Most neurophysiological signals have
a significant amount of energy in very low frequencies. We model this part of the signal with a simple first order SDE of which the covariance function decays exponentially. This process captures some of
the qualitative features of the measured time series, such as roughness
and non-rhythmicity. The model is determined by the following first order
SDE:
\begin{equation}
\frac{d}{dt}\psi(t) = -c\psi(t)+w(t)\,.\label{eq:first order SDE, results}
\end{equation}
The positive number $c$ determines the exponential relaxation of the
process, i.e. how fast its mean decays to zero after a perturbation.
For a compact neuron this is a good model of the sub-threshold membrane
potential under random synaptic inputs~\cite{dayan2001theoretical}. See Fig.~\ref{fig1}D for
some samples of this process. 
\item \textit{Residuals.} Finally, we account for the residuals $\xi(t)$
of our model using a process with temporal covariance that decays
as $e^{-\frac{t^{2}}{2\delta^{2}}}$, where $\delta$ is a small
time constant. This noise is characterized by short-lived temporal
autocorrelations (see Fig.~\ref{fig1}E). As $\delta$ tends to zero,
the process tends to Gaussian white noise. The temporal covariance of this component was not derived from a stochastic differential equation. 
\end{itemize}

\paragraph*{From stochastic differential equations to Gaussian processes regression}

In our dynamical model, the random input
is Gaussian and the dynamics are linear. The linearity implies that
the value of the process at any time point is a linear combination
of the random input at the past time points. As a consequence, because every linear combination
of a set of Gaussian random variables is still Gaussian, the solutions of
the SDEs are Gaussian. The Gaussian Process (GP) distribution
is the generalization of a multivariate Gaussian for infinitely many
degrees of freedom, where the covariance function of the former is
analogous to the covariance matrix of the latter. As a zero-mean multivariate
Gaussian distribution is fully specified by a covariance
matrix, a zero-mean GP $\alpha (t)$ can be completely determined by its covariance
function: 
\begin{equation}
k_{\alpha}(t,t')=\textnormal{cov}(\alpha(t),\alpha(t')) \label{eq:covariance function, results}
\end{equation}
which captures the temporal correlation structure
of the stochastic process $\alpha (t)$. In our case, the covariance function of
the dynamical component $\varphi(t)$, $\chi(t)$ and $\psi(t)$ can be obtained analytically from Eq.~(\ref{eq:oscillator SDE}) and (\ref{eq:first order SDE, results}). This
allows to derive a GP distribution for each linear SDE. Moreover, a sum of independent GPs is again a GP, but now with a covariance
function that is the sum of the covariance functions of each of its
components.  This decomposition of the covariance function is exemplified in Fig.~\ref{fig1}F, which shows the decomposition of the covariance function of a complex signal 
into several component-specific covariance functions, together with
examples of the corresponding dynamic component time series. For 
visual clarity, the second order integrator component has been omitted
from this figure. 

With these GPs as prior distributions, we can use Bayes' theorem for estimating the time course of the dynamic components from the measured time series $y$. In particular, we assume that $y$ is generated by the sum of all dynamic components and corrupted by Gaussian noise $\xi(t)$. The aim is to individually estimate the posterior marginal expectations of $\varphi(t)$, $\chi(t)$ and $\psi(t)$. These marginal expectations are estimates of a dynamic component time course obtained by filtering out from $y$ all the contributions of the other components plus the noise. 

Since both the prior distributions and the observation model are Gaussian, the posterior distribution is itself Gaussian and its marginal expectations can be computed exactly (see Eq.~(\ref{eq:additive covariance, methods}) in Materials and  Methods).

\subsection*{Spatiotemporal GP decomposition}
So far, we have shown how SDE modeling of dynamic
components can be used for analyzing a neural time series through GP regression. Here, we complement
this temporal model by introducing a spatial correlation structure. In this way, we define a full spatiotemporal model. We define the total additive spatiotemporal neural signal as follows:
\[
\rho(\boldsymbol{x},t) = \varphi(\boldsymbol{x},t) + \chi(\boldsymbol{x},t) + \psi(\boldsymbol{x},t)\,, 
\]
where $\boldsymbol{x}$ denotes a cortical location. Strictly speaking, $\rho(\boldsymbol{x},t)$ should be a vector field because the neural electrical activity at each cortical point is modeled as an equivalent current dipole. However, for simplicity, we present the methods for the case in which the dipole orientation is fixed and $\rho(\boldsymbol{x},t)$ can be considered as a scalar field. All formulas for the vector-valued case are given in Appendix~IV. 

\paragraph*{Modeling spatial correlations}
Correlations between different cortical locations can be modeled using a spatial covariance function $s(\boldsymbol{x},\boldsymbol{x}')$. Since the localization of an electric or magnetic source from a sensor array is in general an ill-posed problem, the specification of a prior covariance function is required in order to obtain a unique solution \cite{tarantola2005inverse}. We do not model the spatial correlation structure directly using spatial SDEs. Instead, we impose a certain degree of spatial smoothness, and this is motivated by the fact that fine details of the neural activity cannot be reliably estimated from the MEG or EEG measurements. This procedure has been shown to reduce the localization error and attenuate some of the typical artifacts of source reconstruction \cite{petrov2012harmony,pascual2002functional}.  

Modeling the spatial correlations between measurements of 
neural activity requires a proper definition of distance between cortical
locations. The conventional Euclidean distance is likely to be inappropriate
because cortical gyri can be nearby according to the Euclidean distance
in three-dimensional space, but far apart in terms of the intrinsic cortical
geometry that is determined by the synaptic connectivity between grey matter areas. Surface reconstruction algorithms such as Freesurfer~\cite{fischl2012freesurfer} allow to map each
of the cortical hemispheres onto a sphere in a way that preserves
this intrinsic cortical geometry. Building this spherical
representation, we can make use of the so-called spherical harmonics.
These are basis functions that generalize sines and cosines on the
surface of a sphere and are naturally ordered according to their spatial
frequency. Using the spherical harmonics we define
a spatial covariance function $s(\boldsymbol{x},\boldsymbol{x'})$ between cortical locations, and choose a particular covariance function by discounting
high spatial-frequency harmonics. This operation smooths out the fast-varying neural activity and thereby induces
spatial correlations. This can be interpreted as a low-pass
spatial filter on the cortical surface. The amount
of spatial smoothing is regulated by a smoothing parameter $\upsilon$ and a regularization
parameter $\lambda$, where the former controls the prior spatial correlations and the latter the relative contribution of the prior and the observed spatial correlation (see Eqs.~(\ref{eq:spherical filter, methods}) and (\ref{eq:posterior mean spatiotemporal, methods}) in the Materials and Methods).

\paragraph*{Decomposing spatiotemporal signals using separable covariance functions}

We combine the spatial and temporal model by making a separability assumption, namely we assume that the covariance between $\rho(\boldsymbol{x},t)$ and $\rho(\boldsymbol{x}',t')$  is given by the product $k_\rho(t,t') s(\boldsymbol{x},\boldsymbol{x}')$. Using this spatiotemporal GP prior we compute the marginal expectations of the spatiotemporal dynamic components (see Eq.~(\ref{eq:posterior mean spatiotemporal, methods}) in Materials and Methods). We refer to this approach as spatiotemporal GP decomposition (SGPD).

\subsection*{Estimating the model parameters}
The covariance functions of the dynamic components have parameters that can be directly estimated from the data. Instead of using a full hierarchical model, we estimate the parameters by fitting the total additive covariance function of the model to
the empirical auto-covariance matrix of the measured time series using a least-squares
approach. This procedure allows to infer the parameters of the prior directly from the data, thereby tuning the dynamical model on the specific features of each participant/experimental condition. Specifically, the parameters of the prior are estimated from the data of all trials, and these parameters in turn determine the GP prior distribution that is used for the analysis of the trial-specific data.

The details of the cost function are described in the Materials and Methods
section. Because this optimization problem is not convex, it can have
several local minima. For that reason, we used a gradient-free simulated annealing
procedure \cite{kirkpatrick1984optimization} to find a good approximate solution to the global
optimization problem. 

\subsection*{Analyzing oscillatory amplitude using GP decomposition,
a simulation study }

\paragraph*{ Spectral analysis with temporal GP decomposition}
We now investigate the performance of the temporal GP separation
of dynamic components for the purpose of evaluating modulations of
oscillatory amplitude. Such amplitude modulations have been related
to many cognitive processes. For example, in tasks that require attentional
orienting to some part of the visual field, alpha oscillations are
suppressed over the corresponding brain regions \cite{foxe2011role,kelly2006increases}. Because the spectral content of 
electrophysiological measurements is almost always broadband, when there is
an interest in oscillations, it makes sense to isolate these oscillations from the rest of the measured time series. The resulting procedure involves
a separation of the oscillatory components of interest from the interfering
non-rhythmic components. In the GP decomposition framework, this separation can be achieved by modeling both the oscillatory component $\varphi(t)$ and the interfering processes.
We use the symbol $\boldsymbol{m}_{\varphi|y}$ for the marginal expectation of the process $\varphi(t)$ at the sample points. The average amplitude can be obtained from 
$\boldsymbol{m}_{\varphi|y}$ by calculating its root mean square deviation: 
\begin{equation}
A=\sqrt{\frac{1}{N} \text{\ensuremath{\sum}}_{j}\left(\left[m_{\varphi|y}\right]_{j}-\bar{{m}}\right)^{2}}\label{eq:mean squarred deviation}
\end{equation}
with $\bar{{m}} = \frac{1}{N} \sum_{i}\left[\boldsymbol{m}_{\varphi|y}\right]_{i}$. 

Here, we compared the sensitivity of the GP method with DPSS multitaper spectral estimation \cite{percival1993spectral}, a widely used non-parametric technique.
In the simulation study, the methods had to estimate a simulated
experimental modulation of the amplitude of a 10 Hz oscillatory process.
For each of two conditions, we generated oscillatory time series from
a non-Gaussian oscillatory process. The choice for a non-Gaussian process was motivated by our objective not to bias our evaluation in favor of the GP method.
The oscillatory time series was then corrupted by a first order integrator
and residuals. The simulation design involved 16 levels that
covered an amplitude modulation range from 15\% to 60\% in equidistant
steps. For each level, per experimental condition, we generated 150,000
trials of 2 s. 
The effect size was defined as follows:
\begin{equation}
f=\frac{\langle A_{1}\rangle - \langle A_{2} \rangle}{\text{var}(A)}\,,
\label{eq:effect size, results}
\end{equation}
where $\langle A_{j}\rangle$ is the mean oscillatory amplitude in
the j-th experimental condition and $\mathrm{var}(A)$ is its variance.
Mean and variance were calculated over the trials.

\begin{figure}[]
\centering
    	\includegraphics[] {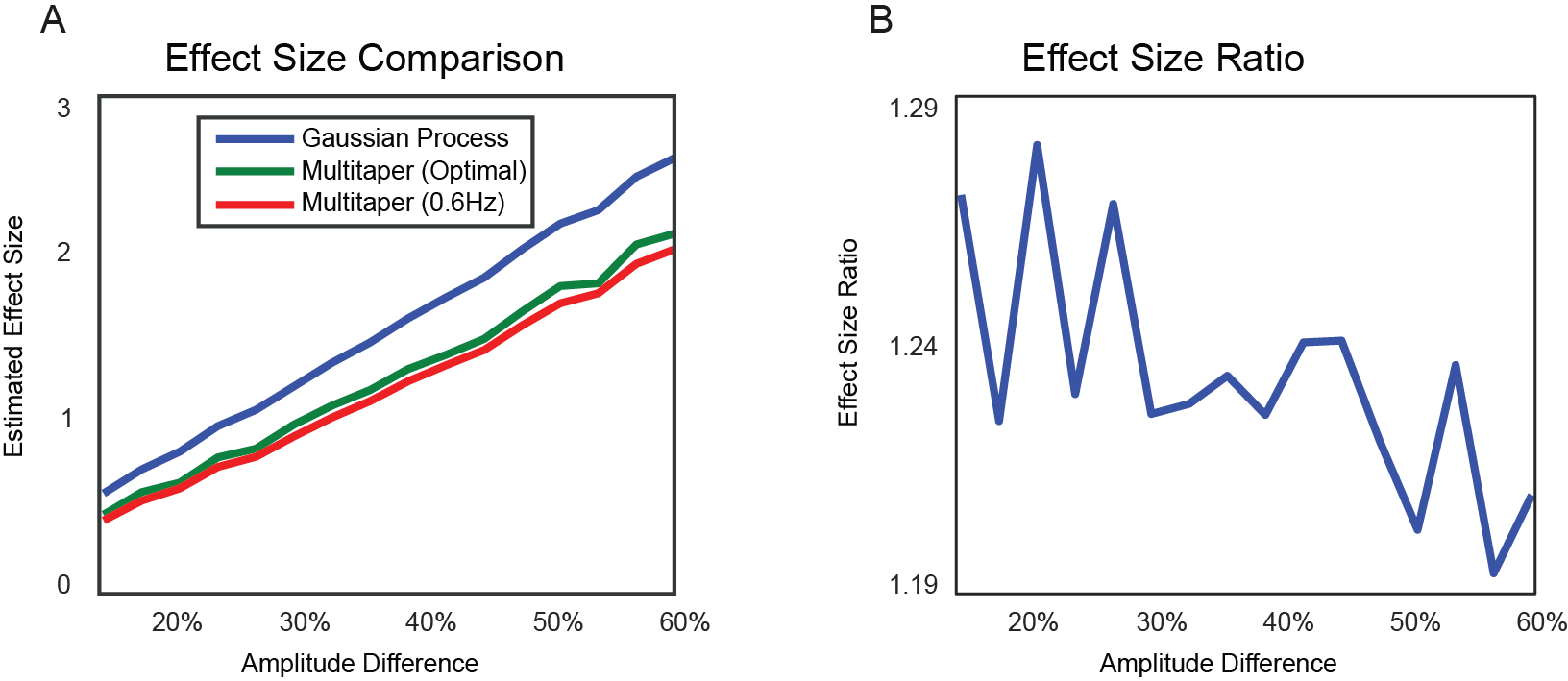}
\caption{\bf Results of the single sensor simulation.}
A) Effect size of temporal GP and DPSS multitaper spectral analysis as function of mean percentage amplitude difference between simulated conditions. The parameters of the temporal GP decomposition (blue line) were estimated from the raw simulated time series. The spectral smoothing of the multitaper method (green line) was chosen for each to maximize the effect size. The red line is the effect size for a multitaper method with constant spectral smoothing of 0.6 Hz. B) Effect size ratio between temporal GP and (optimized) multitaper method as function of the mean amplitude difference between conditions.
\label{fig2}
\end{figure}

The GP method does not have free parameters, since the parameters of the covariance functions are estimated from the data. In contrast,
the spectral smoothing of a multitaper analysis is determined
by the number of tapers, which is a parameter that can be chosen freely.
We selected the number of tapers that maximizes the effect size in
order not to bias the evaluation in favor of the GP method.
In addition, we reported the effect sizes for the multitaper
analysis with a fixed smoothing of 0.6 Hz. Fig.~\ref{fig2}A shows the effect
sizes for the GP and the multitaper method as a function of the true
between-condition amplitude difference. The Gaussian process consistently
outperforms the non-parametric method. Fig.~\ref{fig2}B shows the ratio
between the GP and the optimal multitaper effect size as a function
of the true amplitude difference. Here we can see that the superior
performance is more pronounced when the amplitude difference is smaller, corresponding to a situation with a lower signal-to-noise ratio. 

\paragraph*{SGPD improves accuracy and sharpness of source
reconstruction }

We now investigate how SGPD compares to existing methods with respect to the spatial localization of an oscillatory
amplitude modulation in the presence of noise sources with both spatial
and temporal structure. We compare our method to the Harmony source
reconstruction technique \cite{petrov2012harmony}, which has been shown to outperform
several commonly used linear source reconstruction methods. For this, we set up
a simulation study in which the performance was evaluated by the extent
to which a spatially focal amplitude modulation could be detected.

\begin{figure}[]
\centering
    	\includegraphics[] {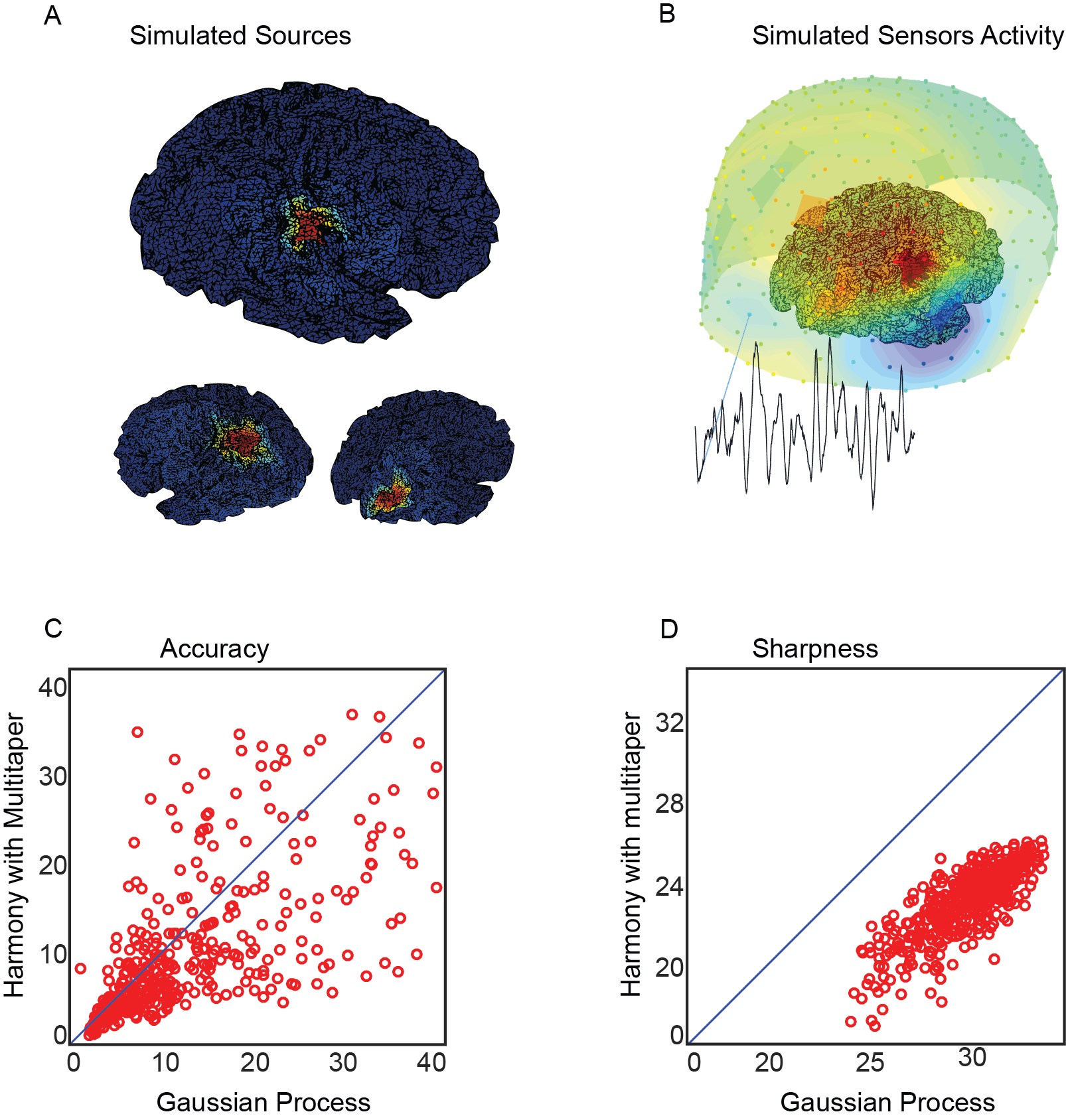}
\caption{\bf Results of the source level simulation.}
A) Spatial maps of the simulated brain sources. The left map shows the spatial extent of the amplitude-modulated source while the two right maps show the interfering sources. The dipole orientation was set to be orthogonal to the mesh surface. B) Visualization of sensor activity as a mixing of the three sources. The dots represent MEG sensors. The color of the dots show the sign (red for positive and blue for negative) together with the magnitudes. The time series was taken from an occipital sensor. C) Scatter plot of the accuracy of SGPD and Harmony. The index was computed by dividing the total reconstructed effect within the amplitude-modulated cortical patch by the sum of total effects in the non-modulated patches. D) Scatter plot of the sharpness of SGPD and Harmony. The sharpness index was obtained by dividing the total reconstructed effect within the amplitude-modulated cortical patch by the total effect elsewhere. For the purpose of visualization, in both scatterplots, we excluded some outliers ($> 5 \times \text{median}$). These outliers arise when the denominator of one of the indices becomes too small. The outliers have been removed from the figure but they were involved in the calculation of the medians for the two methods.
\label{fig3}
\end{figure}

We modeled the brain activity as generated by three cortical patches,
each with a constant spatial profile and a time course generated in
the same way as in the single sensor simulation. The patches had a
radius of approximately one centimeter and were localized in the right
temporal, right occipital, and left parietal cortex (Fig.~\ref{fig3}A). All
three patches exhibited oscillatory activity, but the one in the right
temporal lobe had an amplitude that was modulated by the simulated
conditions. The source activity was projected to the sensors by a
forward model that was obtained using a realistic head model \cite{nolte2003magnetic}. The sensor level activity for the first trial is shown in Fig.~\ref{fig3}B.
The regularization parameter \textgreek{l} of both Harmony and SGPD were identified using leave-one-out
cross validation \cite{stone1974cross}, while the smoothing parameter $\upsilon$ was set
by hand and had the same value of 3 in both models. The spectral smoothing
of the DPSS multitaper spectral estimation was set to 0.6 Hz. The value
was chosen because, on average, this gave the highest effect size of the
amplitude modulation. 

We assessed the quality of the reconstructed
effects using two indices, one for accuracy and one for sharpness.
The accuracy index is obtained by dividing the estimated effect in
the center of the amplitude-modulated patch (more specifically, the sum over the points in a sphere with 1 cm radius) by the maximum of the estimated
effects in the centers of the other two patches (again, by summing
over the points in a sphere). The accuracy index will be high if it
localizes the effect in the right patch but not in the interfering
ones. The sharpness index evaluates how much the effect maps are focused
around the center of the effect. It is computed by dividing the summed
estimated effect in the center of the amplitude-modulated patch by
the summed estimated effect outside that region. Figs.~\ref{fig3}C\&D show
the results of the simulation. Each disc in the scatter plot represents
the outcome of SGPD and Harmony for a single simulation.
The median accuracy and sharpness were respectively 33\% and 28\%
higher for SGPD as compared to the Harmony approach.

\subsection*{Gaussian process analysis of example MEG data }

\begin{figure}[]
\centering
    	\includegraphics[] {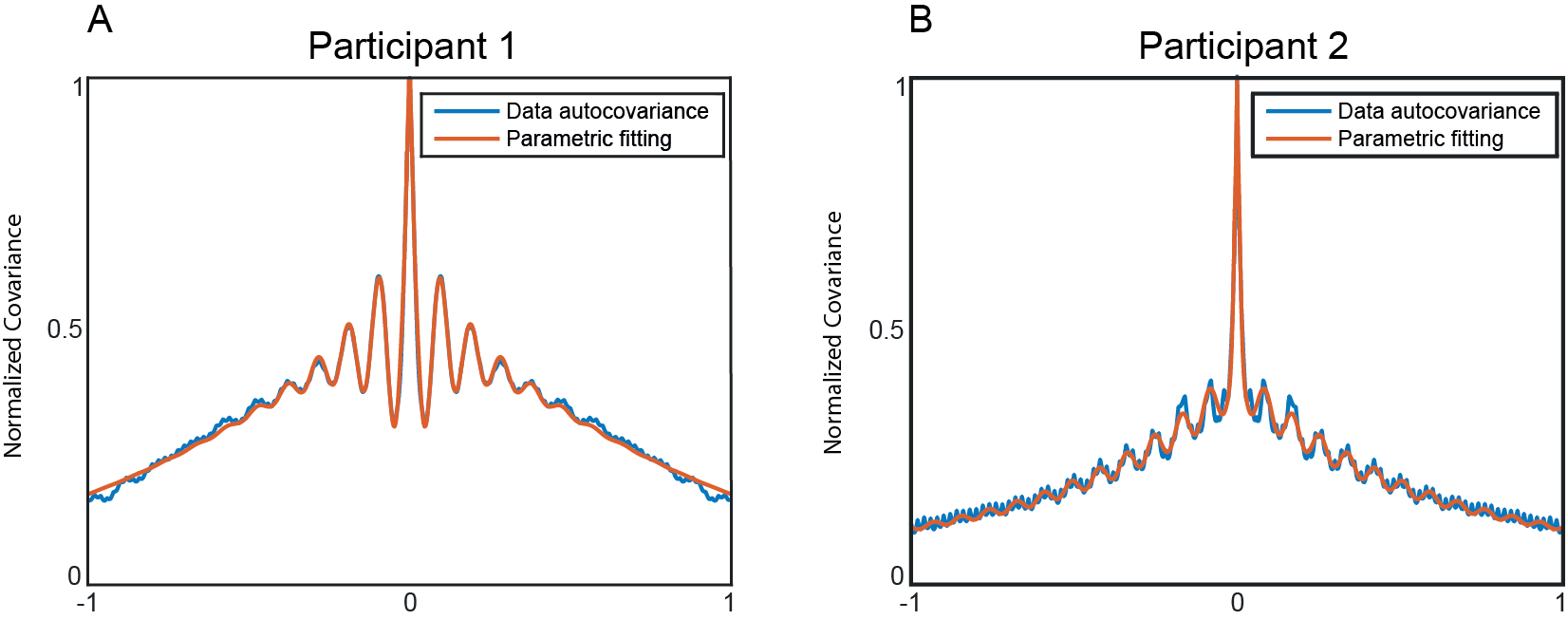}
\caption{\bf Estimation of the model covariance functions.}
Parametric fit of the MEG auto-covariance functions of Participant~1 and Participant~2. The red lines refer to the estimated parametric model and the blue lines reflect the empirical auto-covariance of the measured time series. A single auto-covariance was obtained from the multi-sensor data by performing a principal component analysis and averaging the empirical auto-covariance of the first 50 components, weighted by their variance. The parameters of the model were estimated using a least-squares simulated annealing optimization method. The graphs have been scaled between 0 and 1 by dividing them by the maximum of the individual empirical auto-covariance. 
\label{fig4}
\end{figure}

We tested the temporal GP decomposition on an example MEG dataset that was collected
from 14 participants that performed a somatosensory attention experiment \cite{van2012beyond}. We will use this dataset for different purposes, and start by using it for evaluating the performance of our parameter estimation
algorithm. Fig.~\ref{fig4} shows the empirical auto-covariance functions and the least squares fit for
two participants. To make them comparable,
we normalized these auto-covariance functions by dividing them by their variance. The fitted
auto-covariance functions capture most features of the observed auto-covariance
functions. The comparison shows some individual differences: First, Participant
1 has a higher amplitude alpha signal relative to the other dynamic
components, but the correlation peaks are separated only by about three cycles. Second, the auto-covariance of Participant 2 is
dominated more by a signal component with a high temporal correlation
for nearby points, and the rhythmic alpha component decays much more
slowly. The latter is a signature of a longer phase preservation. 

We quantified the goodness-of-fit as the normalized total absolute
deviation from the model: 
\begin{equation}
g=\frac{\sum_{i,j}|c_{ij}-k(t_{i},t_{j})|}{\sum_{i,j}|c_{ij}|}\,,\label{eq:goodness of fitt, results}
\end{equation}
where $c_{ij}$ is the empirical auto-covariance between $y_{t_{i}}$ and
$y_{t_{j}}$, and $k(t_{i},t_{j})$ is the auto-covariance predicted by
our dynamical model. We evaluated the goodness-of-fit by computing
this deviation measure for each participant. The median goodness-of-fit
was 0.06, meaning that the median deviation from the empirical auto-covariance
was only 6\% of the sum of its absolute values. The goodness-of-fit
for the two example participants one and two in Fig.~\ref{fig4} are 0.04 and 0.02, respectively. 

\begin{figure}[]
\centering
    	\includegraphics[] {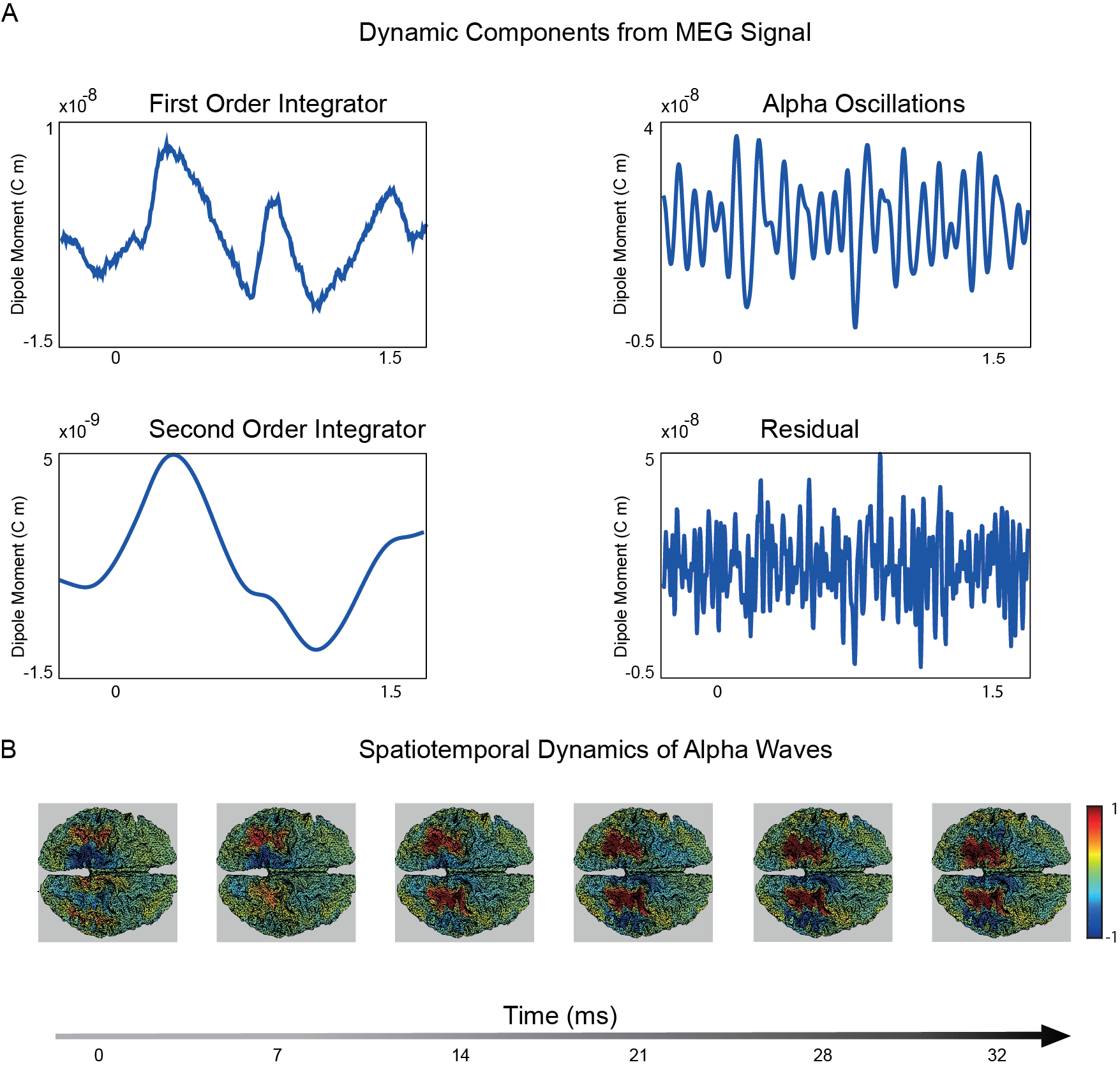}
\caption{\bf Estimated dynamic components.}
Reconstructed source-level neural activity of Participant 1. A) Reconstructed time series of the four dynamic processes localized in a right parietal cortical vertex. B) Reconstructed spatiotemporal dynamics of alpha oscillations along the $x$ axis. This choice of axis is arbitrary and has been chosen solely for visualization purposes. The source-reconstructed activity has been normalized by dividing it by the maximum of the absolute of the spatiotemporal signal.
\label{fig5}
\end{figure}

Next, we inspect the reconstructed spatiotemporal dynamic components obtained
from the resting state MEG signal of Participant 1 (with auto-covariance
as shown in Fig.~\ref{fig4}A), as obtained by SGPD.
Fig.~\ref{fig5}A shows an example of time courses of the dynamic components for an arbitrarily chosen cortical vertex situated in the right parietal cortex. 
The first order integrator time series (upper-left panel) tends to be slow-varying
but also exhibits some fast transitions. The second order integrator (lower-left
panel) is equally slow but smoother. In this participant, the alpha oscillations, as captured by the damped harmonic oscillator,
are quite irregular (upper-right panel), and this is in agreement with
its covariance function (see Fig.~\ref{fig4}A). Finally, the residuals (lower-right
panel) are very irregular, as is expected from the signal\textquoteright s
short-lived temporal correlations. Fig.~\ref{fig5}B shows an example of the
spatiotemporal evolution of alpha oscillations for a period of 32
milliseconds in a resting-state MEG signal. For the purpose of visualization,
we only show the value of the dipole along an arbitrary axis. The
pattern in the left hemisphere has a wavefront that propagates through the parietal cortex. Conversely, the alpha signal in
the right hemisphere is more stationary. 

\subsection*{Attention-induced spatiotemporal dynamics of oscillatory amplitude}

Next, we applied the SGPD source reconstruction method to the
example MEG data that were collected in a cued tactile detection experiment. Identifying the neurophysiological
mechanisms underlying attentional orienting is an active area of investigation
in cognitive neuroscience~\cite{foxe2011role, jensen2010shaping, van2011orienting, van2012beyond}. Such mechanisms could involve neural activity of which the spatial distribution varies over time (i.e., neural activity with dynamic spatial patterns), and GP source
reconstruction turns out to be highly suited for identifying such
activity, as we will demonstrate now. 

In the cued tactile detection
experiment an auditory stimulus (high or low pitch pure tone) cued
the location (left or right hand) of a near-threshold tactile stimulus
in one-third of the trials. This cue was presented 1.5 s before
the target. The remaining two-thirds of the trials were uncued. In
the following, we compare the pre-target interval between the cued
and the uncued conditions in terms of how the alpha
amplitude modulation develops over time. In the analysis, we made
use of the fact that the experiment involved two recording sessions,
separated by a break. We explored the data of the first session in
search for some pattern, and then used the data of the second session
to statistically test for the presence of this pattern. Thus,
the spatiotemporal details of the null hypothesis of this statistical
test were determined by the data of the first session, and we used
the data of the second session to test it.

\begin{figure}[]
\centering
    	\includegraphics[] {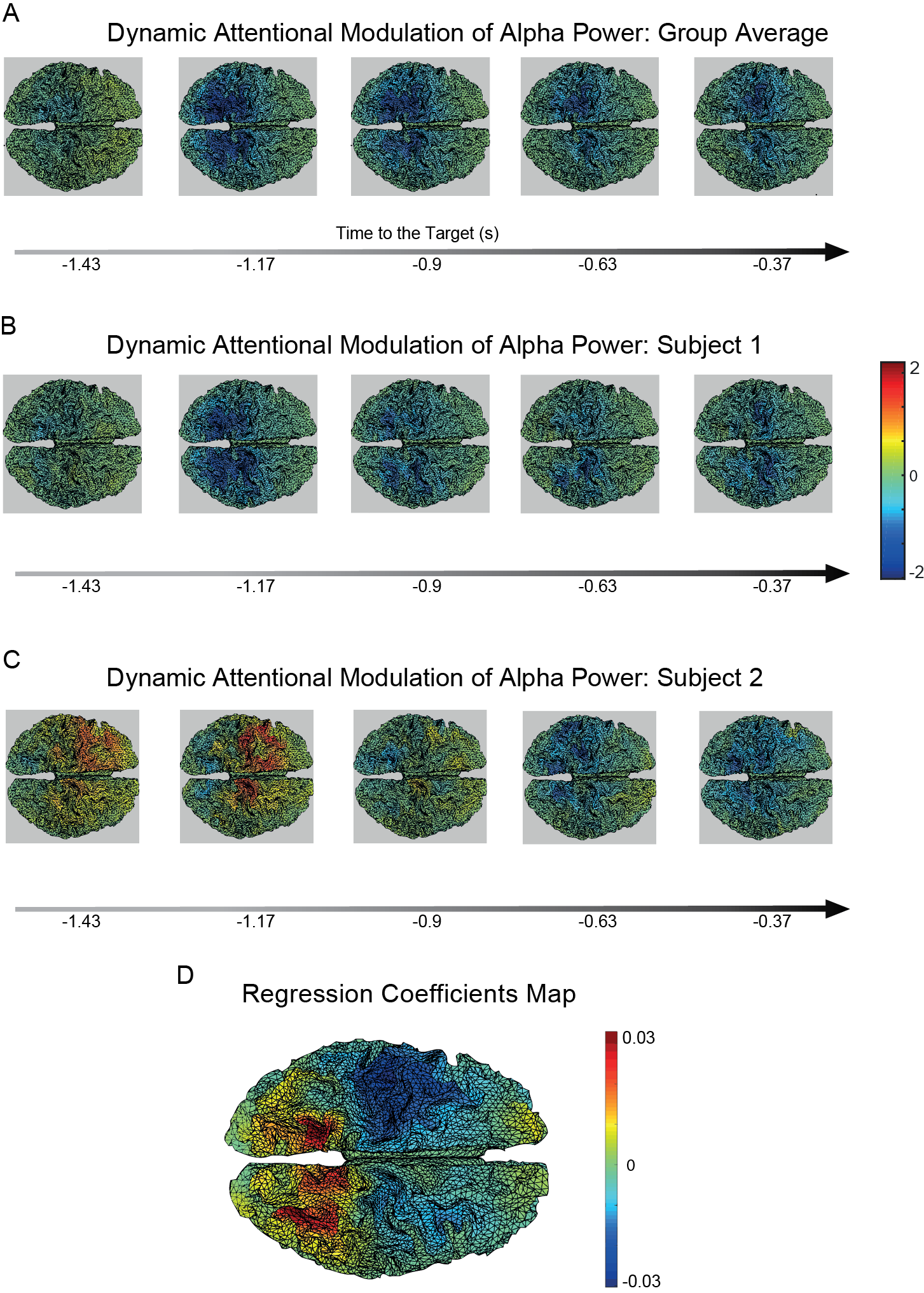}
\caption{\bf Caudal-to-rostral progression of alpha amplitude attentional modulation.}
A) Group average of alpha amplitude attentional modulation as function of time. B,C) Alpha amplitude attentional modulation for participants 1 and 2, respectively. D) Spatial map obtained by computing the slope of the average alpha difference between cued and non-cued conditions as a function of time for each cortical vertex.
\label{fig6}
\end{figure}

Figure~\ref{fig6}A shows the group-averaged
alpha amplitude modulation as a function of time. An amplitude suppression
for the cued relative to the uncued condition originates bilaterally
in the parietal cortex and gradually progresses caudal to rostral
until it reaches the sensorimotor cortices. The time axes are expressed in terms of the distance to the target. Similar patterns can be
seen in individual participants (see Fig.~\ref{fig6}B\&C for representative
participants 1 and 2). Participant 1 has a suppressive profile that
is almost indistinguishable from the group average. On the other hand, participant 2 shows an early enhancement of sensorimotor alpha
power accompanied by a parietal suppression, and the latter then propagates
forward until it reaches the sensorimotor areas. Thus, in the grand
average and in most of the participants, there is a clear caudal-to-rostral
progression in the attention-induced alpha amplitude suppression.
We characterized this progression by constructing cortical maps of
the linear dependence (slope) between latency and amplitude modulation.
The group average of the slope maps for the first session is shown
in Fig.~\ref{fig6}D. This figure shows that the posterior part of the brain
has positive slopes, reflecting the fact that the effect tended to
become less negative over time. Conversely, the sensorimotor regions
have positive slopes, reflecting the fact that the effect tended to
become more negative over time. 

To evaluate the reliability of this pattern, we build on the reasoning that, if this pattern in the slope map is due to chance,
then it must be uncorrelated with the slope map for the second session.
To evaluate this, for every participant, we calculated the dot product between the normalized slope maps for the two sessions and tested whether the average dot product was different from zero.

The one-sample t-test showed that the
effect was significantly different from zero ($p<0.05$), supporting
the claim that the caudal-to-rostral progression in the attention-induced
alpha amplitude suppression is genuine. Thus, we have shown that,
during the attentional preparation following the cue, the alpha modulation
progresses from the parietal to the sensorimotor cortex. 

\section*{Discussion}

In this paper, we introduced a new signal decomposition technique that incorporates
explicit dynamical models of neural activity. We showed how dynamical models can be constructed and integrated into a Bayesian statistical analysis framework based on
GP regression. The resulting statistical model can be used for decomposing
the measured time series into a set of temporal or spatiotemporal
dynamic components that reflect different aspects of the neural signal.
We validated our method using simulations and real
MEG data. The simulations demonstrate that the use of the dynamical
signal and noise model improves on the current state-of-the-art (non-parametric spectral estimation) with respect to the identification of amplitude
modulations between experimental conditions. Different from non-parametric spectral estimation, our method first separates the oscillatory signal from the interfering components on the basis of the autocorrelation structures of both, and only in the second step calculates the amplitude of the signal of interest.
A spatiotemporal version of the decomposition method was obtained by decomposing the neural processes in spherical harmonics.
Our simulations show that, in the presence of spatially and temporally correlated noise, spatiotemporal GP decomposition localizes amplitude modulations more accurately than a related method that does not make use of the temporal decomposition of the signal of interest \cite{petrov2012harmony}.
Lastly, using the spatiotemporal decomposition on real MEG data from a somatosensory detection task, we demonstrated its usefulness by identifying an intriguing anterior-to-posterior propagation in the attention-induced suppression of oscillatory alpha power.

\subsection*{Generality, limitations, and robustness}

Although we used a specific set of SDEs, the method is fully general in that it can be applied to any linearized model of neuronal activity. Therefore, it establishes a valuable connection between data analysis
and theoretical modeling of neural phenomena. For example, neural
masses models and neural field equations (see, e.g. \cite{deco2008dynamic}) can be
linearized around their fixed points and the resulting SDEs form the basis for a GP analysis that extracts the theoretically defined components. Furthermore, the GP decomposition method could be used as an analytically solvable starting point for the statistical analysis of non-linear and non-Gaussian phenomena through methods such as perturbative expansion, where the initial linear Gaussian model is corrected by non-linear terms that come from the Taylor expansion of the non-linear couplings between the neural activity at different spatiotemporal points~\cite{dickman2003path}. 

The method's limitations pertain to the model's assumptions of linearity and its specific spatial correlation structure. Specifically, the linear SDEs cannot account for the complex non-linear effects that are found in both experimental \cite{freyer2009bistability, osorio2010epileptic} and modeling work \cite{golos2015multistability, ghosh2008noise, breakspear2006unifying}. In addition, the assumed homegeneous spatial correlation structure solely depends on the distance between cortical locations and therefore does not account for the rich connectivity structure of the brain \cite{greicius2009resting, stam2004functional, sporns2005human, van2011rich}. Nevertheless, the method has some robustness against the violations of the underlying assumptions. This robustness follows from the fact that the model specifies the prior distribution but does not constrain the marginal expectations to have a specific parametric form. The temporal prior affects the estimation of a dynamic component to a degree that depends on the ratio between its variance and the cumulative variance of all other components. Specifically, the smaller the prior variance of a component relative to the combined variance of all the others, the more the pattern in the prior covariance matrix will affect the posterior. Since we estimate all these prior variances directly from the measured time series, our method is able to reconstruct complex non-linear effects in components that have a relatively high SNR while it tends to ''linearize'' components with low SNR. As a consequence, the more pronounced the non-linear effects in the observed signal, the more these will be reflected in the posterior, gradually dominating the linear structure imposed by the prior. Importantly, because our temporal prior is based on a larger data set, it will be adequate, on average, over all epochs while still allowing strong components in individual epochs to dominate the results. 

The situation is similar but not identical for our spatial prior. Contrary to our temporal prior, this spatial prior is not derived from an empirically fitted dynamical model but on the basis of our prior belief that source configurations with high spatial frequencies are unlikely to be reliably estimated from MEG measurements. Since the problem of reconstructing source activity from MEG measurements is generally ill-posed, the choice of the spatial prior will bias the inference even for very high SNR. Nevertheless, it has been shown that the discounting of high spatial frequencies leads to reduced localization error and more interpretable results \cite{petrov2012harmony}. 

\subsection*{Connections with other methods}

The ideas behind the GP decomposition derive from a series of recent developments
in machine learning, connecting GP regression to stochastic dynamics
\cite{sarkka2012infinite,solin2013infinite}. The approach is closely connected with many methods in
several areas of statistical data-analysis. We will now review some of these links, focusing on methods that are commonly used in neuroscience.

\paragraph*{Spectral analysis}

Throughout the paper, we showed that the GP decomposition can be profitably used
to estimate amplitude modulations in an oscillatory signal, which
is an important application of spectral analysis. There are two classes of spectral analysis methods: parametric and non-parametric \cite{percival1993spectral}. 
Non-parametric methods mostly rely on the discrete Fourier transform applied to a tapered signal, as for example in DPSS multitaper spectral estimation \cite{percival1993spectral, bronez1992performance}. These methods are non-parametric because they do not explicitly model
the process that generates the signal. 

Parametric methods {\em do\/} depend on an explicit model, and typically this is an autoregressive (AR) model \cite{muthuswamy1998spectral,pace1998spatiotemporal}. AR models are closely related to GPs as they are typically formulated as discrete-time Gaussian processes driven by stochastic {\em difference \/} equations. In this sense, the GP prior distributions used in this paper are continuous-time versions of an AR process. However, the usual AR approach to spectral
estimation is different from our approach. AR models are usually parametrized in terms of a series of matrices whose entries
(the so-called model coefficients) describe the statistical dependencies within and between the channels of a multivariate signal. The coefficients that describe the statistical dependencies within a univariate signal are related to the inverse of the impulse response function in our approach (see Materials and Methods for a description
of the impulse response function). Spectral analysis based on AR models has the disadvantage that a very large number of parameters may have to be estimated. Specifically, a spatiotemporal AR model for multi-sensor EEG/MEG data can easily have
hundreds of model coefficients that must be estimated from the measured time series.
This great flexibility in the analytic form of the AR model is required
as the spectrum is directly obtained from the model coefficients.

Compared to spatiotemporal AR modeling, the GP decomposition model
is much more constrained by the underlying theory, having an explicit
additive structure with few parameters for each dynamic component.
The rigidity of the model is compensated by the fact that the oscillatory
amplitude is not obtained from the fitted model covariance function.
Instead, it is computed from the marginal expectation of the oscillatory component, which is obtained
by applying Bayes' rule. Therefore, while the prior has a parametric
form, the posterior mean obtained from the GP decomposition method is actually non-parametric. 

\paragraph*{Signal decomposition}

In the neuroscience literature, the most widely used signal separation
techniques are the blind source separation methods known as principal component analysis (PCA) and
independent component analysis (ICA), together with their extensions \cite{comon1994independent,tipping1999probabilistic,zou2006sparse,scholkopf1998nonlinear,kolda2009tensor, van2012phase,van2015uncovering}. These methods rely on the statistical properties
of multi-sensor data (maximum variance for PCA and statistical
independence for ICA) and produce components whose associated signals
are linear combinations of the sensor-level signals. Specifically,
these statistical properties pertain to the resulting component-level signals. Importantly, whereas GP decomposition depends on a specific model of the neural signal, neither PCA nor ICA makes use of
prior knowledge of the component-level signals. Also, both
PCA and ICA require multi-sensor data, whereas GP decomposition
can be applied to a single time series. 

It is important to note that GP decomposition is not a
tool for separating statistically independent or uncorrelated components. Instead, its goal is to decompose the measured
signal into several processes characterized by different autocorrelation structures. Hence, the method does not discriminate between two independent processes generated by two
sources with the same dynamics, such as a frontal and an occipital alpha
oscillation. Therefore, the GP decomposition is
complementary to blind source separation. In fact, the latter can
be used to extract interesting temporal and spatiotemporal patterns
from the dynamic components obtained from GP decomposition.

\paragraph*{Source reconstruction}

A general framework for GP source analysis has recently been introduced~\cite{solin2016gprec}. In this work, the authors show that several well-known source reconstruction methods are special cases of GP regression with appropriate covariance functions. In particular, the spatial filter of techniques such as minimum norm estimation~\cite{hamalainen1994interpreting} and exact Loreta \cite{pascual2002functional} are obtained as a
discretization of a spatial GP analysis with an appropriate spatial
covariance function. The authors also introduced a general framework for GP spatiotemporal analysis using separable covariance functions designed to localize averaged neural activity (e.g. evoked fields). This GP spatiotemporal source reconstruction is formally similar to several other spatiotemporal source reconstruction
methods~\cite{darvas2001spatio,friston2008multiple,trujillo2008bayesian,dannhauer2013spatio}. 

Our approach improves on these works by using informed temporal covariance functions that explicitly model the temporal dynamics of the ongoing neural signal. The additive structure of the temporal covariance function allows to individually source localize signal components with specific dynamic properties. In particular, the spatial configuration of these components are analyzed in the spherical harmonics domain, as this greatly reduces the dimensionality of the source space. As shown in the Materials and Methods section, the resulting spatial filter is closely related to the Harmony source reconstruction method~\cite{petrov2012harmony}. 

\subsection*{Benefits of GP Decomposition for Cognitive Neuroscience}
In two simulation studies, we showed that the GP decomposition improves the quantification and source localization of oscillatory amplitude modulations. These improvements are particularly noticeable in the presence of the kind of temporally and spatially structured noise that is ubiquitous in neural measurements. Additionally, the method is particularly suited for data-driven exploration of complex spatiotemporal data as it decomposes the signal into a series of more interpretable dynamic components.

As a demostration, we used the SGDP to investigate the modulation of alpha oscillations associated with attentional preparation to a tactile stimulus. Several previous works demonstrated that alpha amplitude is reduced prior to a predicted stimulus~\cite{van2012beyond, van2011orienting, foxe2011role}. These amplitude modulations have been associated to modality specific preparatory regulations of the sensory cortices~\cite{van2008prestimulus,klimesch2007eeg,van2011orienting,volberg2009eeg,rohenkohl2011alpha}. While the attentional role of alpha oscillations in the primary sensory cortices is well established, it is still unclear how this generalizes to supramodal areas. Although the parietal cortex is known to play a role in the top-down control of attention~\cite{culham2001neuroimaging, corbetta2002control}, parietal alpha oscillations have typically been considered as closely related to the visual system~\cite{foxe2011role}.
 
The involvement of the parietal cortex in the somatosensory detection task went unnoticed in the first analysis of the data that have been reanalyzed in the present paper~\cite{van2012beyond}. In our new analysis, we used the SGDP to more effectively explore the data, looking for interesting spatiotemporal effects. This led to the identification of a suppression of alpha amplitude that originates from the parietal cortex and then propagates to the somatosensory regions. This effect turned out to be statistically robust when tested in a second independent dataset that was collected in the same experiment. The results suggest a hierarchical organization of the reconfiguration of alpha amplitude following an attentional cue. In particular, the initial reduction of parietal alpha amplitude could reflect the activation of a supramodal attentional network that paves the way for later sensorimotor-specific cortical reconfiguration.

While we mainly restricted our attention to the analysis of alpha oscillations, we believe that the GP decomposition can be useful for the study of other neural oscillations as well as non-rhythmic components. Several experimental tasks are related to effects in multiple dynamic components. For example, perception of naturalistic videos induces modulations in several frequency bands \cite{betti2013natural}. Studying the interplay between these differential modulations requires an appropriate decomposition of the measured signals that can be effectively performed using GP decomposition.

\subsection*{Conclusions }
Our dynamic decomposition method starts from a precise mathematical model of the dynamics of the neural fields. The formalism of GP regression allows translation of linear stochastic dynamics into a well-defined Bayesian prior distribution. In this way, the method establishes a connection between mathematical modeling and data analysis of neural phenomena. On the one hand, the experimentalist and the data-analyst can benefit from the method as it allows to isolate the dynamic components of interest from the interfering noise. These components are interpretable and visualizable, and their study can lead to the identification of new temporal and spatiotemporal neural phenomena that are relevant for human cognition. On the other hand, the theorist can use this formalism for obtaining a probabilistic formulation of dynamical models, thereby relating them to the experimental data. 

\section*{Materials and Methods}
In this section we will explain the mathematical underpinnings of the GP decomposition. Following the lines of the Results section, the exposition begins from the connection between SDEs and Gaussian processes and continues with the exposition of the temporal and spatiotemporal GP decomposition. In order to improve the readability and to not overshadow the main ideas, we left some technical derivations to the appendices.

\subsection*{From SDEs to GPs}
At the core of our method is the connection between Gaussian processes and SDEs. This connection leads to the definition of the covariance functions of the dynamic components that will be used for determining the prior of the GP regression. In the Results section, we introduced the SDE (Eq.~(\ref{eq:oscillator SDE}))
\[\frac{d^{2}}{dt^{2}}\varphi(t)+b\frac{d}{dt}\varphi(t)=-\omega_{0}^{2}\varphi(t)+w(t)\]
to model an oscillatory signal. In fact, this SDE can be interpreted as a damped harmonic oscillator when $b<\sqrt{2\omega_0^2}$. As initial conditions, we set $\varphi(-\infty) = \frac{d\varphi}{dt}(-\infty) = 0$. This choice implies that the (deterministic) effects of the initial conditions are negligible. Given these initial conditions, the solution of Eq.~(\ref{eq:oscillator SDE}) is fully specified by the random input $w(t)$ that follows a temporally uncorrelated normal distribution. Since the equation is linear, the solution, given a particular instantiation of $w(t)$, can be obtained by convolving $w(t)$ with the impulse response function of the SDE (see Appendix~I for more details):
\begin{equation}
\varphi(t) = \int_{-\infty}^{\infty} G_\varphi(t - s) w(s)ds.\label{eq:green function, methods}
\end{equation}

Intuitively, the impulse response function $G_\varphi(t)$ determines the response of the system to a localized unit-amplitude input. Consequently, Eq.~(\ref{eq:green function, methods}) states that the process $\varphi(t)$ is generated by the infinite superposition of responses to $w(t)$ at every time point. This proves that the resulting stochastic process $\varphi(t)$ is Gaussian, since it is a linear mixture of Gaussian random variables.

The impulse response function of Eq.~(\ref{eq:oscillator SDE}) is
\begin{equation}
G_\varphi (t)= \vartheta(t) e^{-b/2 t}  \sin \omega t, \label{eq:Green oscillator, methods}
\end{equation}
where $\vartheta(t)$ is a function equal to zero for $t<0$ and 1 otherwise. This function assures that the response cannot precede the input impulse. From this formula, we see that the system responds to an impulse by oscillating at frequency $\omega = \sqrt{\omega_0^2 - 1/4b^2}$ and with an amplitude that decays exponentially with time scale $b/2$. The covariance function of the process $\varphi(t)$ can be determined from its impulse response function using Eq.~(\ref{eq:deriv. prior cov ||, methods}) (see Appendix~I) and is given by
\begin{equation}
k_\varphi(t_i,t_j) = k_\varphi(\tau) = \frac{\sigma_\varphi^2} {2b} e^{-b/2 |\tau|} \left( \cos \omega \tau + \frac {b} {\omega} \sin \omega |\tau| \right). \label{eq:Cov oscillator, methods}
\end{equation}
where $\tau$ denotes the time difference $t_i - t_j$. In the case of the second order integrator, the parameter $\omega_0$ is smaller than $b/2$ and the system is overdamped. In this case, the response to an impulse is not oscillatory, the response initially rises and then decays to zero with time scale $b/2$. This behavior is determined by the impulse response function
\begin{equation}
G_\chi (t)= \vartheta(t) e^{-b/2 t}  \sinh z t \label{eq:Green integrator, methods}
\end{equation}
in which $z$ is equal to $\sqrt{1/4 b^2-\omega_0^2}$. 
The covariance function is given by
\begin{equation}
k_\chi(\tau) = \frac{\sigma_\chi^2} {2b} e^{-b/2 |\tau|} \left( \cosh z \tau + \frac {b} {z} \sinh z |\tau| \right). \label{eq:Cov integrator, methods}
\end{equation}
Finally, the first order integrator (Eq.~(\ref{eq:first order SDE, results}))
\[\ \frac{d}{dt}\psi(t)-c\psi(t)+w(t) \]
has a discontinuous impulse response function that decays exponentially:
\begin{equation}
G_{\psi} (t)= \vartheta(t) e^{-c t}\,. \label{eq:Green first order, methods}
\end{equation}

The discontinuity of the impulse response at $t = 0$ implies that the process is not differentiable as it reacts very abruptly to the external input. The covariance function of this process is given by:
\begin{equation}
k_{\psi}(\tau) = \frac{\sigma_{\psi}^2} {2c} e^{-c |\tau|}\,. \label{eq:Cov first order, methods}
\end{equation}

\paragraph{Covariance function for the residuals}

The stochastic differential equations are meant to capture the most important (linear) qualitative features of the neural signal. Nevertheless, the real underlying neural dynamics are much more complex than can be captured by any simple model. Empirically, we found that the residuals of our model have short-lived temporal correlations. We decided to account for these correlations by introducing a residuals process $\xi(t)$ with covariance function
\begin{equation}
k_{\xi}(\tau) = \sigma_{\xi}^2  e^{- \frac {\tau^2} {2 \delta^2}} \label{eq:Cov res, methods}
\end{equation}
in which the small time constant $\delta$ is the signal's characteristic time scale and $\sigma_{\xi}$ is its standard deviation. This covariance function is commonly called the squared exponential and is one of the most used in the machine learning literature \cite{rasmussen2006gaussian}. As $k_{\xi}(\tau)$ decays to zero much faster than our SDE-derived covariance functions for $\tau$ tending to  $\infty$, this covariance function is appropriate for modeling short-lived temporal correlations.

\subsection*{Analysing neural signals using Gaussian process regression }

In this section, we show how to estimate the value of a dynamic component such as $\varphi (t)$ in the set of sample points $t_1,…,t_N$ using GP regression. To this end, it is convenient to collect all the components  other than $\varphi (t)$ in a total residuals process $\zeta (t) = \chi (t) + \psi (t) + \xi (t)$. In fact, in this context, they jointly have the role of interfering noise. The vector of data points $\boldsymbol{y}$ is assumed to be a sum of the signal of interest and the noise:
\begin{equation}
    y_j = \varphi (t_j) + \zeta (t_j)\,. \label{eq:Bayesian model likelihood I, methods}
\end{equation}

In order to estimate the values of $\varphi (t)$ using Bayes' theorem we need to specify a prior distribution over the space of continuous-time signals. In the previous sections, we saw how to construct such probability distributions from linear SDEs. In particular, we found that those distributions were GPs with covariance functions that can be analytically obtained from the impulse response function of the SDEs. These prior distributions can be summarized in the following way:
\begin{align}
     \varphi (t) &\sim GP(0,k_\varphi (t_1, t_2) ) \\
     \zeta (t) &\sim GP(0,k_{\zeta} (t_1, t_2) ) \nonumber  \label{eq:Bayesian model prior I, methods}
\end{align}
where the symbol $\sim$ indicates that the random variable on the left-hand side follows the distribution on the right-hand side and $GP(\mu(t),k(t_1,t_2))$ denotes a GP with mean function $\mu(t)$ and covariance function $k(t_1,t_2)$. Note that, in this functional notation, expressions such as $\mu(t)$ and $k(t_1,t_2)$ denote whole functions rather than just the values of these functions at specific time points.

We will now derive the marginal expectation of $\varphi(t)$ under the posterior distribution. Since we are interested in the values of $\varphi(t)$ at sample points $t_1,\ldots,t_N$, it is convenient to introduce the vector $\boldsymbol{\varphi}$ defined by the entries $\varphi_j = \varphi(t_j )$. 
Any marginal distribution of a GP for a finite set of sample points is a multivariate Gaussian whose covariance matrix is obtained by evaluating the covariance function at every pair of time points:
\begin{equation}
[K_\varphi]_{ij} = k_\varphi(t_i, t_j). \label{eq:Cov matrix, methods}
\end{equation}
Using Bayes' theorem and integrating out the total residual $\zeta(t)$, we can now write the marginal posterior of $\boldsymbol{\varphi}$ as
\begin{equation}
p(\boldsymbol{\varphi} \mid \boldsymbol{y} ) \propto \int p(\boldsymbol{y}\mid \boldsymbol{\varphi}, \boldsymbol{\zeta}) p(\boldsymbol{\zeta}) d\boldsymbol{\zeta}\ p(\boldsymbol{\varphi}) = N(\boldsymbol{y} \mid \boldsymbol{\varphi},K_{\zeta} )N(\boldsymbol{\varphi} \mid 0,K_\varphi) \label{eq:Bayes theorem, methods}
\end{equation}
in which $K_{\zeta}$ is the temporal covariance matrix of $\zeta (t)$.  As a product of two Gaussian densities, the posterior density is a Gaussian distribution itself. The parameters of the posterior can be found by writing the prior and the likelihood in canonical form. From this form, it is easy to show that the posterior marginal expectation is given by the vector $\boldsymbol{m}_{\varphi|y}$ (see~\cite{rasmussen2006gaussian} for more details about this derivation):
\begin{equation}
\boldsymbol{m}_{\varphi|y} = K_{\varphi} (K_{\varphi} + K_{\zeta})^{-1} \boldsymbol{y}. \label{eq:temporal posterior mean, methods}
\end{equation}

Furthermore, if we assume that $\chi (t)$, $\psi (t)$ and $\xi (t)$ are independent, the noise covariance matrix reduces to
\begin{equation}
K_{\zeta} = K_{\chi} + K_{\psi} + K_{\xi}. \label{eq:additive covariance, methods}
\end{equation}

\subsection*{GP analysis of spatiotemporal signals}

In the following, we show how to generalize GP decomposition to the spatiotemporal setting. This requires the construction of a source model and the definition of an appropriate prior covariance between cortical locations. In fact, the problem of localizing brain activity from MEG or EEG sensors becomes solvable once we introduce prior spatial correlations by defining a spatial covariance $s(\boldsymbol{x}_i,\boldsymbol{x}_j )$ between every pair of cortical locations $\boldsymbol{x}_i$ and $\boldsymbol{x}_j$. In this paper, we construct $s(\boldsymbol{x}_i,\boldsymbol{x}_j)$ by discounting high spatial frequencies in the spherical harmonics domain, thereby limiting our reconstruction to spatial scales that can be reliably estimated from the sensor measurements. However, prior to the definition of the covariance function, we need to specify a model of the geometry of the head and the brain cortex.

\paragraph*{The source model }
In order to define a source model, we construct a triangular mesh of the cortex from a structural MRI scan using Freesurfer \cite{fischl2012freesurfer}. The cortical boundary is morphed into a spherical hull in a way that maximally preserves the intrinsic geometry of the cortex. This allows to parameterize the surface $C$ using the spherical coordinates $\alpha$  and $\theta$, respectively azimuth and elevation. For notational simplicity, we collect the spherical coordinates into the coordinate pair $\boldsymbol{x}=(\alpha,\theta)$ that refers to a spatial location in the cortex. Furthermore, we denote the finite set of $M$ points in the mesh as $\mathcal{X} = \{\boldsymbol{x}_1,\ldots,\boldsymbol{x}_M\}$. 

We define our source model as a vector field of current dipoles on the cortical surface. We first consider GP source reconstruction of the total neural activity $\vec{\rho}(\boldsymbol{x}, t)$, without differentiating between spatiotemporal dynamic components such as $\vec{\varphi}(\boldsymbol{x}, t)$,$\vec{\chi}(\boldsymbol{x}, t)$ and $\vec{\psi}(\boldsymbol{x}, t)$. The vector field $\vec{\rho}(\boldsymbol{x}, t)$ is characterized by the three Cartesian coordinates $\rho_1 (\boldsymbol{x},t)$, $\rho_2 (\boldsymbol{x},t)$, and $\rho_3 (\boldsymbol{x},t)$. In all the analyses contained in this paper, we estimate the full vector field. However, since we do not assume any prior correlations between the dipole coordinates, in the following we will simplify the notation by describing the source decomposition method for a dipole field $\rho (\boldsymbol{x},t) \vec{v}(\boldsymbol{x})$, where the unit-length vector field $\vec{v}(\boldsymbol{x})$ of dipole orientations is assumed to be known. Appendix~IV explains how to adapt all the formulas to the vector-valued case using matrices with a block diagonal form. 

\paragraph*{Spatial Gaussian processes source reconstruction in the spherical harmonics domain}

The linearity of the electromagnetic field allows to model the spatiotemporal data matrix $Y$ as the result of a linear operator  acting on the neural activity $\rho (\boldsymbol{x},t)$ \cite{nolte2003magnetic}:
\begin{equation}
Y_{ij} = \int_C \mathcal{L}_i (\boldsymbol{x}) \rho (\boldsymbol{x}, t_j) d \boldsymbol{x}~,
\label{eq:leadifield, methods}
\end{equation}
in which the component $\mathcal{L}_i (\boldsymbol{x})$ describes the effect of a source located at $\boldsymbol{x}$ on the $i$-th sensor. Note that $\mathcal{L}_i (\boldsymbol{x})$ implicitly depends on the orientation $\vec{v}(\boldsymbol{x})$ since different dipole orientations generate different sensor measurements. We refer to $\mathcal{L}_i (\boldsymbol{x})$ as the forward model relative to the $i$-th sensor, note that this is a function of the spatial location on the cortical surface.

\begin{figure}[]
\centering
    	\includegraphics[] {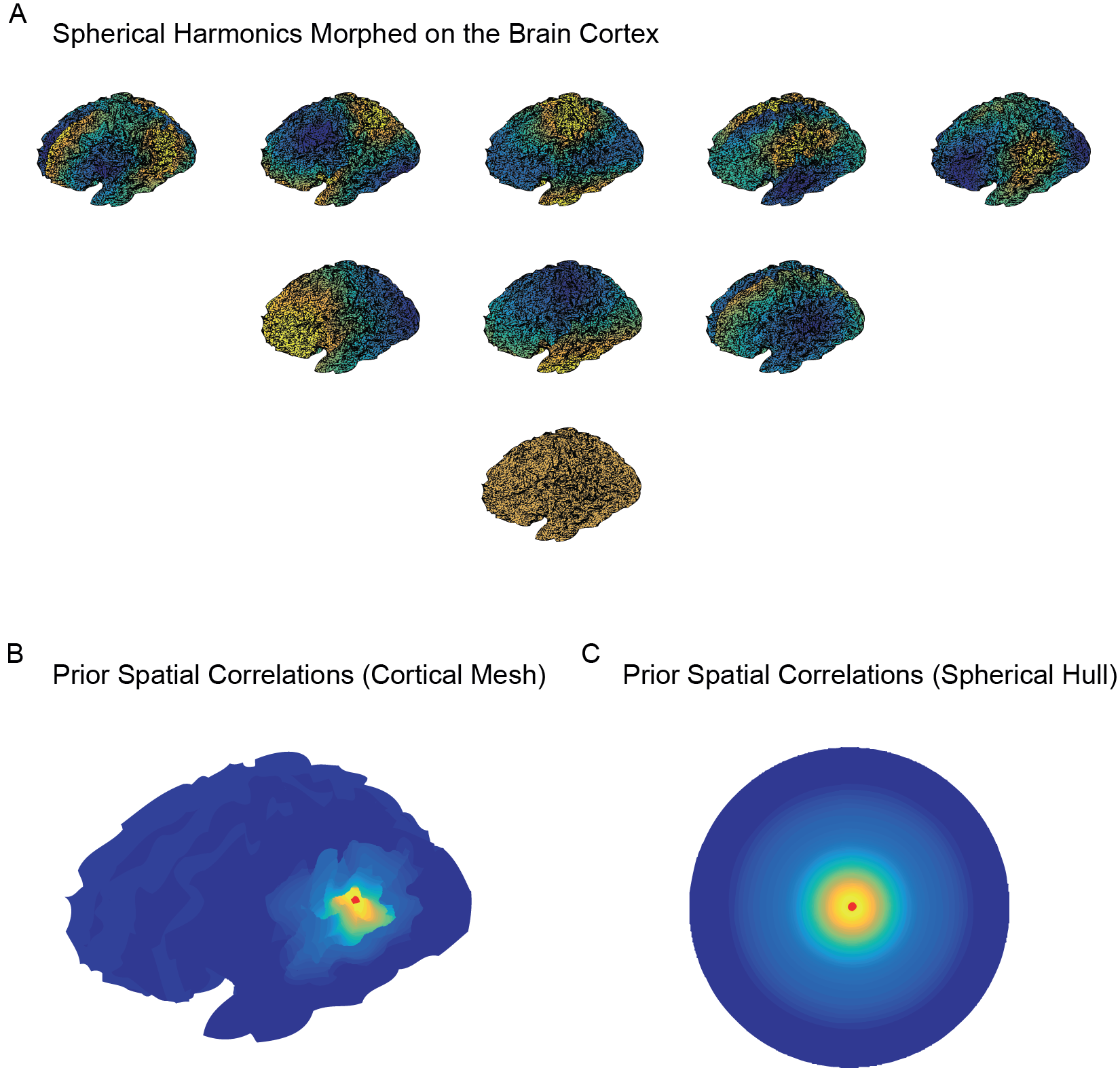}
\caption{\bf Spherical harmonics and covariance functions.}
Visualization of the spherical harmonics morphed onto the cortex and the resulting spatial correlation structure. A) Example of spherical harmonics on the brain cortex for frequency numbers from 0 to 2. For each frequency number $l$ there are $2 l + 1$ harmonics with ''phase'' number $m$ ranging from $-l$ to $l$. As clear from the picture, the spatial frequency increases as a function of the frequency number. In all our analyses we truncated the harmonic expansion after the 11th frequency number.  B,C) Prior correlation structure induced by Eq.~(\ref{eq:spatial covariance, methods}). Panel B shows the prior correlations on the cortical surface from a cortical point identified by a red dot. Panel C shows the same function on the spherical hull. The spatial correlations are determined by the frequency discount function $f(l)$; here we used the same smoothing parameters as all analyses in the paper: $k =2$ and $\upsilon = 3$. 
\label{fig7}
\end{figure}

In this section, we ignore the prior temporal correlations induced by the temporal covariance functions, i.e. we implicitly assume a prior for $\rho(\boldsymbol{x}, t)$ that is temporally white. In a GP regression setting, the spatial smoothing can be implemented by using a spatially homogeneous covariance function, i.e. a covariance function that only depends on the cortical distance between the sources. To define this covariance function, we make use of the so-called spherical Fourier transform. Whereas the ordinary Fourier transform decomposes signals into sinusoidal waves, the spherical Fourier transform decomposes spatial configurations defined over a sphere into the spherical harmonics $\mathcal{H}_l^m(\boldsymbol{x})$. These basis functions are characterized by a spatial frequency number $l$ and a ''spatial phase'' number $m$. Fig.~\ref{fig7}A  shows the spherical harmonics corresponding to the first three spatial frequencies morphed on the cortical surface. For notational convenience, we assign an arbitrary linear indexing to each $(l,m)$ couple that henceforth will be denoted as $(l_k,m_k)$. It is convenient to represent the neural activity $\rho(\boldsymbol{x},t)$ in the spherical harmonics domain. Specifically, we will use the symbol $\tilde{\rho} (l_k,m_k; t)$ to denote the Fourier coefficient of the spherical harmonic indexed by $(l_k,m_k)$ (see Eq.~(\ref{eq:direct Fourier, methods}) in Appendix~II).

We assume that the spherical Fourier coefficients $\tilde{\rho} (l_k,m_k; t)$ are independent Gaussian random variables. Under this assumption, we just need to define the prior variance of the coefficients $\tilde{\rho} (l_k,m_k; t)$. Since we aim to reduce the effect of noise with high spatial frequencies, we define these prior variances using a frequency damping function $f(l_k)$ that monotonically decreases as a function of the spatial frequency number $l_k$. This effectively discounts high spatial frequencies and therefore can be seen as a spherical low-pass filter. The variance of the spherical Fourier coefficients is given by the following variance function
\begin{equation}
\tilde{s} (l_k,m_k; t) = f(l_k), \label{eq:fourier variance, methods}
\end{equation}
where, as damping function, we use a spherical version of the truncated Butterworth low-pass filter:
\begin{equation}
f(l_k) = 
\begin{cases}
{\Big( 1 + (\frac{l_k}{\upsilon})^{2k} \Big) }^{-1/2} & \text{for} \ l_k \leq L\\
0 & \text{for} \ l_k > L 
\end{cases} \label{eq:spherical filter, methods}
\end{equation}
with smoothing parameter $\upsilon$, order $k$, and cut-off frequency $L$. This filter has been shown to have good properties in the spatial domain \cite{devaraju2012performance}. Note that, under the covariance function defined by Eq.~(\ref{eq:fourier variance, methods}) and (\ref{eq:spherical filter, methods}), the spherical Fourier coefficients with frequency number larger than $L$ have zero variance and are therefore irrelevant. Although the analysis is carried out in the spherical harmonics domain, it is informative to be able to visualize the covariance function in the spatial domain. By applying the inverse spherical Fourier transform, the function $s(\boldsymbol{x}_i,\boldsymbol{x}_j)$ can be explicitly obtained as follows:
\begin{equation}
s(\boldsymbol{x}_i,\boldsymbol{x}_j) = \sum_{l,m} \mathcal{H}_l^m(\boldsymbol{x}_i) \mathcal{H}_l^m(\boldsymbol{x}_j) f(l) \,. \label{eq:spatial covariance, methods}
\end{equation}
Figs.~\ref{fig7}B and \ref{fig7}C show the correlations induced by our spatial covariance function.

In order to formulate the spatial GP regression in the spherical harmonics domain, we rewrite the integral in Eq.~(\ref{eq:leadifield, methods}) using the inverse spherical Fourier transform (see Eq.~(\ref{eq:inverse Fourier, methods}) in Appendix~II) and interchanging the order of summation and integration:
\begin{equation}
Y_{ij} =  \int_C \mathcal{L}_i (\boldsymbol{x}) \Bigg({\sum_{k} \tilde{\rho} (l_k,m_k; t_j) \mathcal{H}_{l_k}^{m_k}(\boldsymbol{x})}\Bigg) d\boldsymbol{x} = \sum_{k} \tilde{\mathcal{L}}_i (l_k,m_k)\tilde{\rho} (l_k,m_k; t_j) , \label{eq:fourier leadifield, methods}
\end{equation}
where 
\begin{equation}
\tilde{\mathcal{L}}_i (l_k,m_k) = \int_C \mathcal{L}_i (\boldsymbol{x}) \mathcal{H}_{l_k}^{m_k}(\boldsymbol{x}) d\boldsymbol{x}
\end{equation}
is the spherical Fourier transform of $\mathcal{L}_i (\boldsymbol{x})$. Therefore, the spherical Fourier transform converts the forward model (which is a function of the cortical location) from the spatial to the spherical harmonics domain. We can simplify Eq.~(\ref{eq:fourier leadifield, methods}) by organizing the spherical Fourier coefficients $\tilde{\rho} (l_k,m_k; t_j)$ in the matrix $\tilde{R}$, whose element $\tilde{R}_{kj}$ is $\tilde{\rho} (l_k,m_k; t_j)$. Analogously, the spherical Fourier transform of the forward model can be arranged in a matrix $\Lambda$ with elements $\Lambda_{ik} = \tilde{\mathcal{L}}_i (l_k,m_k)$. Using this notation, we can write the observation model for the spatiotemporal data matrix $Y$ in a compact way: 
\begin{equation}
Y = \Lambda \tilde{R} + \boldsymbol{\xi}~,
\label{eq:spatial observation model, methods}
\end{equation}
where $\boldsymbol{\xi}$ are Gaussian residuals with spatial covariance matrix $\Sigma$. 

We can now combine this observation model with the spherical harmonics domain spatial GP prior, as determined by the variance function given by Eq.~(\ref{eq:fourier variance, methods}), and from this we obtain the posterior of the neural activity $\tilde{R}$ given the measured signal $Y$. Because the spatial process is Gaussian, the prior distribution of the spherical Fourier coefficients is normal and, because we assumed that the spherical Fourier coefficients are independent, their covariance matrix $D$ is diagonal with entries specified by the variance function $D_{kk} = f(l_k)$ (see Eq.~(\ref{eq:fourier variance, methods})). Alltogether, the prior and the observation model specify a Gaussian linear regression. The posterior expectation of the regression coeffcients $\tilde{R}$ can be shown to be \cite{tarantola2005inverse}:
\begin{equation}
M_{\tilde{R}|Y} = D \Lambda^{T} ( \Lambda D \Lambda^{T} + \Sigma)^{-1} Y. \label{eq:spatial fourier posterior mean, methods}
\end{equation}
In this formula, $\Lambda D \Lambda^{T}$ is the sensor level covariance matrix induced by the spatially smooth brain activity and $\Sigma$ is the residual covariance matrix of the sensors. This expression can be recast in terms of the original cortical locations $\mathcal{X}$  using the inverse spherical Fourier transform (Eq.~(\ref{eq:inverse Fourier, methods})). In matrix form, this can be written as
\begin{equation}
M_{R|Y} = H M_{\tilde{\rho}|Y},
\label{eq:spatial posterior mean, methods}
\end{equation}
where the matrix $H$ is obtained by evaluating the spherical harmonics at the discrete spatial grid-points $\mathcal{X}$:
\begin{equation}
H_{lk} = \mathcal{H}_{l_k}^{m_k}(\boldsymbol{x}_l)\,. \label{eq:harmonic matrix, methods}
\end{equation}
This formula gives the Harmony source reconstruction solution as presented in~\cite{petrov2012harmony}. We can reformulate this expression by introducing the Harmony spatial filter 
\begin{equation}
P = H D \Lambda^{T} ( \Lambda D \Lambda^{T} + \Sigma)^{-1}.
\label{eq:spatial filter, methods}
\end{equation}
Using this matrix, the posterior expectation of the neural activity at the cortical locations $\mathcal{X}$ can be written as follows:
\begin{equation}
M_{R|Y} = P Y.
\label{eq:spatial posterior mean II, methods}
\end{equation}

\paragraph*{Spatiotemporal GP decomposition}

The temporal and spatial GP regression can be combined by assigning a temporal covariance function to each spherical Fourier coefficient. In other words, we model the time series of each coefficient as an independent temporal Gaussian process. These processes have the same prior temporal correlation structure as specified in our additive temporal model. However, as in the spatial model, their prior variance is discounted as a function of the spatial frequency $l_k$. Using functional notation, this can be written as follows:
\begin{equation}
\tilde{\rho} (l,m;t) \sim GP(0,f(l) k_{\rho} (t_1, t_2)). \label{eq:spatiotemporal process, methods}
\end{equation}
Considering the prior distributions of the processes $\tilde{\rho} (l,m;t)$ at the sample points, the matrix-valued random variable $\tilde{R}$, when vectorized, follows a multivariate Gaussian distribution with covariance matrix $K_{\rho} \otimes D$, where $\otimes$ denotes the Kronecker product (see Appendix~III). This Kronecker product form follows from the fact that the covariance function of $\tilde{\rho} (l,m;t)$ is the product of a spatial and a temporal part. Multivariate Gaussian distributions with this Kronecker structure can be more compactly reformulated as a matrix normal distribution (see \cite{dawid1981some}):
\begin{equation}
\tilde{R} \sim MN(0, D, K_{\rho})\,, \label{eq:matrix normal prior, methods}
\end{equation}
where the matrix parameters $D$ and $K_{\rho}$ determine the covariance structure across, respectively, the spherical harmonics and time.  

We define a spatiotemporal observation model in which the residuals have a spatiotemporal covariance structure of the form $K_{\xi} \otimes (\Lambda D \Lambda^T)$. This implies that the spatial covariance matrix of the residuals (previously denoted as $\Sigma$) has the form $\Lambda D \Lambda^T$. Thus, it is assumed that the residuals have the same spatial covariance as the brain activity of interest 
(see Eq.~(\ref{eq:spatial fourier posterior mean, methods})) but a different temporal covariance.
Hence, $\xi(\boldsymbol{x},t)$ should be interpreted as brain noise \cite{mivsic2010brain}. This assumption greatly simplifies the derivation of the posterior distribution. Under this observation model, the probability distribution of the spatiotemporal data matrix can be written as follows:
\begin{equation}
Y \sim MN(\Lambda \tilde{R}, \Lambda D \Lambda^T, K_{\xi})~. \label{eq:matrix normal likelihood, methods}
\end{equation}

The posterior expectation for this model can be obtained using the properties of Kronecker product matrices. This derivation is slightly technical and is reported in Appendix~III. In this derivation, to enhance numerical stability, we introduce a Tikhonov regularization parameter $\lambda$. This allows us to deal with the fact that the matrix $\Lambda D \Lambda^{T}$ (which must be inverted), is usually close to singular for an MEG or EEG forward model. The resulting posterior expectation is the following:
\begin{equation}
M_{\tilde{R}|Y} = D \Lambda^T ( \Lambda D \Lambda^{T} + \lambda I)^{-1} Y (K_{\rho} + K_\xi)^{-1} K_{\rho} \,. 
\label{eq:posterior mean spatiotemporal, methods}
\end{equation}
Besides regularizing the matrix inversion, the $\lambda I$ term contributes to filtering out the spatially non-structured observation noise. This is consistent with the fact that the regularization matrix replaces the noise spatial covariance matrix in Eq.~(\ref{eq:spatial fourier posterior mean, methods}) and, being diagonal, corresponds to spatially white noise. In the spatial domain, Eq.~(\ref{eq:posterior mean spatiotemporal, methods}) becomes:
\begin{equation}
M_{R|Y} = P Y (K_{\rho} + K_\xi)^{-1} K_{\rho}\,. \label{eq:posterior mean spatiotemporal II, methods}
\end{equation}
Therefore, the spatiotemporal expectation is obtained by applying the Harmony spatial filter (with $\Sigma = \lambda I$) to the expectation of the temporal model given by Eq.~(\ref{eq:temporal posterior mean, methods}). We can now apply this to the situation in which we want to estimate some component of interest, such as $\varphi(\boldsymbol{x},t)$, in the presence of other components $\zeta(\boldsymbol{x},t) = \chi(\boldsymbol{x},t) + \psi(\boldsymbol{x},t) + \xi(\boldsymbol{x},t)$. In analogy with Eq.~(\ref{eq:temporal posterior mean, methods}), the marginal expectation of the spatiotemporal component $\varphi(\boldsymbol{x},t)$ is given by
\begin{equation}
M_{\Phi|Y} = P Y (K_{\varphi} + K_{\zeta})^{-1} K_{\varphi}\,, \label{eq:posterior mean spatiotemporal III, methods}
\end{equation}
where $K_{\zeta}$ is the temporal covariance matrix of $\zeta(\boldsymbol{x},t) = \chi(\boldsymbol{x},t) + \psi(\boldsymbol{x},t) + \xi(\boldsymbol{x},t)$. This formula allows to individually reconstruct the dynamic components.

\subsection*{Estimating the model parameters}

We estimate the parameters of the covariance functions from all the data of each participant using an empirical Bayes method. This produces a prior distribution that is both informed by the participant-specific signal dynamics and flexible enough to account for the variability across different epochs. Specifically, given $K$ trials, the parameters are estimated from the empirical autocovariance matrix $S$ of the total measured time series: 
\begin{equation}
S = \sum_{k=1}^K Y_{k} Y_{k}^T \label{eq:empiric covariance, methods}
\end{equation}
where $Y_{k}$ denotes the demeaned (mean-subtracted) spatiotemporal data matrix of an experimental trial $k$. For notational convenience, we organize all the parameters of the model covariance function in the vector $\boldsymbol{\vartheta}$. Furthermore, we make the dependence on the parameters explicit by denoting the total covariance function of the total additive model as 
\begin{equation}
k_{\rho}(t,t'; \boldsymbol{\vartheta}) = k_{\varphi}(t,t'; \boldsymbol{\vartheta}) + k_{\chi}(t,t'; \boldsymbol{\vartheta}) + k_{\psi}(t,t'; \boldsymbol{\vartheta}) + k_{\xi}(t,t'; \boldsymbol{\vartheta}). \label{eq:total covariance, methods}
\end{equation}

As the objective function to be minimized, we use the sum of the squared deviations of the measured time series' auto-covariance from the covariance function of our model: 
\begin{equation}
C(\boldsymbol{\vartheta}) = \sum_{i,j} \bigg(S_{ij} - k_{\rho}(t_i, t_j ; \boldsymbol{\vartheta})\bigg)^2 \label{eq:cost function, methods}
\end{equation}
This objective function is, in general, multimodal and requires the use of a robust optimization technique. Gradient-based methods can be unstable since they can easily lead to sub-optimal local-minima. For that reason we used a gradient-free simulated annealing strategy. The details of the simulated annealing algorithm are described in \cite{kirkpatrick1984optimization}. As proposal distribution we used 
\begin{equation}
p(\vartheta^{(k+1)}_j) = t(\vartheta^{(k+1)}_j|\vartheta^{(k)}_j,\gamma_j,1)\,,
\label{eq:proposal distribution, methods}
\end{equation}
where $t(x|a,b,c)$ denotes a univariate Student's t-distribution over $x$ with mean $a$, scale $b$ and $c$ degrees of freedom. We chose this distribution because the samples can span several order of magnitudes, thereby allowing both a quick convergence to the low cost region and an effective fine tuning at the final stages. 
We used the following annealing schedule:
\begin{equation}
T(n + 1) = 0.8 \cdot T(n)\,, 
\label{eq:temperature schedule, methods}
\end{equation}
where $T(0)$ was initialized at $10$ and the algorithm stopped when the temperature was smaller than $10^{-8}$. 

We estimated all the temporal parameters of the model. Specifically, the estimated parameters were the following: (a) the alpha frequency $\omega = \sqrt{\omega_0^2 - 1/4b^2}$, phase decay $\beta_{\varphi} = 1/2b_{\varphi}$, and amplitude $\mathcal{A}_{\varphi} = \sigma_{\varphi} / \sqrt{2b_{\varphi}}$, (b) the second order integrator parameters $z$, $\beta_{\chi} = 1/2b_{\chi}$, and its amplitude $\mathcal{A}_{\chi} = \sigma_{\chi} / \sqrt{2 b_{\chi}}$, (c) the first order integrator decay constant $c$ and its amplitude $\mathcal{A}_{\psi} = \sigma_{\psi} / \sqrt{2 b_{\psi}}$, and (d) the residual's time scale $\delta$, and standard deviation $\sigma_{\xi}$. The parameters were initialized at plausible values (e.g. 10 Hz for the oscillator frequency) and were constrained to stay within realistic intervals ( 6--15 Hz for alpha frequency, positive for $\beta_{\varphi}$, $\beta_{\chi}$, $c$, $\delta$ and all the amplitudes).

\subsection*{Details of the simulation studies}
\paragraph*{Single sensor simulation} 
The simulation was composed of two ''experimental'' conditions that differed only with respect to the mean of the oscillatory amplitude. The simulation design involved 16 levels, with amplitude differences ranging from 15\% to 60\%. For each level, we generated 150,000 trials per experimental condition, which gave us very reliable estimates of the effect size. The trials were 2 s long.
In order to not give an unfair advantage to our method (which is based on Gaussian processes), the trial time series were generated as a non-Gaussian random process according to the following formula:
\begin{equation}
y(t)=  \sqrt{a^2(t)+1}  \cos(\omega(t)t+ \gamma)+\xi(t)+\psi(t) \,. \label{eq:syntethic signal I, methods}
\end{equation}
The random initial phase $\gamma$ 
in this formula was drawn from a uniform distribution, and the functions $a(t)$ and $\omega(t)$ are Gaussian processes with a squared exponential covariance function (see Eq.~(\ref{eq:Cov res, methods})). The mean of the angular frequency $\omega(t)$ was equal to $2\pi \cdot 10$ (the typical frequency of alpha oscillations) for both experimental conditions. The noise processes $\xi(t)$ and $\psi(t)$ were generated by, respectively, a first order integrator and a residual Gaussian process (see Eq.~(\ref{eq:Cov res, methods}) and~(\ref{eq:Cov first order, methods})). 
We used the temporal GP decomposition to extract the oscillatory component from the simulated time series. The effect sizes were quantified as the between-condition differences between the trial-averaged amplitudes divided by the across-trials standard deviation of the amplitudes.

We compared the sensitivity of the GP decomposition with the non-parametric spectral estimation using DPSS multitaper spectral analysis as described in \cite{percival1993spectral}. For every trial, the mean oscillatory amplitude was obtained by averaging over the amplitude estimates for the orthogonal tapers. In this method, the number of tapers is a free parameter that determines the degree of spectral smoothing. For each cell of the simulation design, we chose the number of tapers that maximizes the effect size. This selection procedure is biased in favor of the multitaper method since it tends to overfit the data and therefore produces larger effect sizes.

\paragraph*{Source level simulation}
A template cortical surface mesh was created using Freesurfer \cite{fischl2012freesurfer}, down-sampled using the MNE toolbox \cite{gramfort2014mne}, and aligned to a template MEG sensor configuration.
We ran 500 trials, each involving two conditions that differed only with respect to the oscillatory amplitude of one cortical location. Sources were generated at three locations in the brain: one in the right parieto-temporal, one in the right occipital and one in the left parietal cortex. For each trial and condition, we generated three time series with the same temporal structure as those generated in the single sensor simulation study. The three time series were localized in cortical mesh with a spatial profile that is proportional to a Fisher-von Mises distribution. These spatial profiles can model a localized patch of activity. The dipole orientation was set to be orthogonal to the mesh surface. While all patches of activity contained the oscillatory component, only one patch involved an amplitude modulation between the two experimental conditions, and this was set at 20\%. 
The activity was projected to the MEG sensors using a forward model obtained from a realistic head model \cite{nolte2003magnetic}. 
The effect was computed for each cortical vertex as the difference in average oscillatory amplitude between the two conditions. 

The oscillatory signal was first reconstructed at each cortical vertex using the spatiotemporal GP decomposition. Next, as in the simulation study for the single sensor, the GP estimate of average oscillatory amplitude was obtained as the standard deviation of the estimate of the oscillatory component.
We compared the spatiotemporal GP decomposition with the Harmony source reconstruction of the estimated cross-spectral density matrix. Using the DPSS multitaper spectral analysis, we first estimated the sensor-level cross-spectral density matrix $F$. Next, we projected this matrix to the source level by sandwiching it between the Harmony spatial filters (see Eq.~(\ref{eq:spatial posterior mean, methods})): 
$
F_H = P F P^T
$.
The source level amplitude is obtained by taking the square root of the diagonal elements of $F_H$. The spectral smoothing was kept fixed at 0.6 Hz since we found this value to be optimal given the simulation parameters. 

\subsection*{Details of the application to an MEG study on anticipatory spatial attention}

\paragraph*{Participants and data collection}
We tested the spatiotemporal GP source reconstruction method on a cued tactile detection experiment in which the magneto-encephalogram (MEG) was recorded [26]. The study was conducted in accordance with the Declaration of Helsinki and approved by the local ethics committee (CMO Regio Arnhem-Nijmegen). Informed written consent was obtained from all participants. Fourteen healthy participants (5 male; 22–49 yr) participated in the study. 
The MEG system (CTF MEG; MISL, Coquitlam, British Columbia, Canada) had 273 axial gradiometers and was located in a magnetically shielded room. The head position was determined by localization coils fixed to anatomic landmarks (nasion and ears). The data were low-pass filtered (300-Hz cutoff), digitized at 1,200 Hz and stored for offline analysis. 

\paragraph*{Experimental design}
The experiment was a tactile detection task in which the location and timing of the targets were either cued or not. A short auditory stimulus (50 ms, white noise) was presented together with an electrotactile stimulus (0.5-ms electric pulse close to threshold intensity) in half of the trials. In the other half the auditory stimulus was presented alone.  Participants were asked to indicate if a tactile stimulus was presented. In one-third of the trials, an auditory cue (150 ms, pure tone) informed the participants about the timing and the location at which the tactile stimulus might occur. In particular, the target auditory signal was always presented 1.5 s after the cue. Two independent sessions were collected for each participant. More details can be found in \cite{van2012beyond}.

\paragraph*{MEG preprocessing}
Third-order synthetic gradients were used to attenuate the environmental noise \cite{vrba2001signal}. In addition, extra-cerebral physiological sources such as heartbeat and eye movements were detected using independent component analysis \cite{comon1994independent} and regressed out from the signal prior to the spatiotemporal GP decomposition. 

\paragraph*{Details of the GP spatiotemporal data analysis}
We started the GP analysis by learning the parameters of the additive dynamical model for each individual participant using the simulated annealing method. To reduce the contribution of low-amplitude noise, we estimated this matrix from the first 50 principal components of the total empirical temporal cross-covariance matrix averaged over all channels.
A template cortical surface mesh was created using Freesurfer \cite{fischl2012freesurfer}, downsampled using the MNE toolbox \cite{gramfort2014mne}, and aligned to the MEG sensors using the measured head position.
The Tikhonov regularization parameter $\lambda$ was identified for each participant using leave-one-out cross-validation \cite{stone1974cross}. The spatial smoothing parameters $k$ and $\upsilon$ were set to, respectively, 2 and 3.
The spatiotemporal GP decomposition was applied to 1.8 s long segments, starting ten milliseconds before the presentation of the cue and ending ten milliseconds after the target stimulus. The alpha amplitude envelope $A(t,x)$ was obtained for all cortical vertices and dipole directions by performing a Hilbert transform on the estimated alpha signal and taking the absolute value of the resulting analytic signal [61]. For each cortical location, the total amplitude was obtained by summing the amplitude envelopes for the three independent dipole directions $\varphi_1 (\boldsymbol{x},t)$, $\varphi_2 (\boldsymbol{x},t)$, and $\varphi_3 (\boldsymbol{x},t)$. The individual topographic maps of the attention-induced alpha amplitude suppression were obtained by computing the mean amplitude difference between cued and non-cued trials, separately for each vertex and time point. These individual maps were then averaged across participants, again for each vertex and time point.

\paragraph*{Statistical analysis}

For each cortical vertex, the dynamic effect was quantified as the rate of change of the attention-induced alpha amplitude suppression as a function of elapsed time from cue onset. Specifically, we used linear regression to estimate the slope of the relation between attention-induced alpha amplitude suppression and time. We did this separately for every vertex. 
The cortical maps of regression coefficients were constructed from the first experimental session of every participant and then averaged across participants. This map was subsequently used as data-driven hypothesis which was tested using the data from the second session. As a test statistic, we used the dot product between the individual regression coefficients maps, computed from the second sessions, and the group-level map. Under the null hypothesis that the group-level map is not systematic (i.e., is driven by noise only), the expected value of this test statistic is zero. Therefore we tested this null hypothesis using a one-sample t-test. 

\section*{Appendix I: Covariance functions defined by linear SDEs} \label{sec: appendix I}

Consider a general linear SDE of the form
\begin{equation}
\sum_{k}^K c_{k} \frac {d^k\alpha(t)} {dt^k} = w(t)\label{eq:general SDE, methods}
\end{equation}
where the coefficients $c_{k}$ are chosen in a way to have stable solutions. An important tool for analyzing a linear differential equation is the impulse response function $G(t)$. This function is defined as the response of the system to a unit-amplitude impulse $\delta(t)$:
\begin{equation}
\sum_{k}^K c_{k} \frac {d^k G(t)} {dt^k} = \delta(t)\label{eq:green function definition, methods}
\end{equation}
Using the impulse response function, a solution of the linear SDE driven by an arbitrary random input $w(t)$ can be written as follows: 
\begin{equation}
\alpha(t) = \int_{-\infty}^{\infty} G (t- s)  w(s)ds.\label{eq:green function, appendix}
\end{equation}
This means that the stochastic process $\alpha(t)$ is an infinite linear superposition of responses to the random uncorrelated input $w(s)$.  

Using Eq.~(\ref{eq:green function, appendix}) we can derive the mean and covariance function of $\alpha(t)$. The mean function is defined as 
\begin{equation}
m_{\alpha}(t)= \langle \alpha (t) \rangle,\label{eq:prior expectation,appendix}
\end{equation}
where the triangular brackets $\langle \cdot \rangle$ denote the expectation with respect of the distribution of the random input $w(s)$. Using (\ref{eq:green function, appendix}) in (\ref{eq:prior expectation,appendix}), we obtain:
\begin{equation}
m_{\alpha}(t) = \int_{-\infty}^{\infty} G (t - s)  \langle w(s) \rangle ds =0. \label{eq:deriv. prior mean, appendix}
\end{equation}
Here, we used the fact that the order of expectation and integration can be interchanged and that the expectation of the white noise process is equal to zero. Analogously, we can obtain the covariance function as follows:
\begin{equation}
k_{\alpha} (t,t' )= \langle \alpha(t)\alpha(t') \rangle = \int_{-\infty}^{\infty} \int_{-\infty}^{\infty} G (t- s)  G (t'- s') \langle w(s)w(s' \rangle ds ds'. \label{eq:deriv. prior cov |, appendix}
\end{equation} 
Since $w(s)$ is white, its covariance $\langle w(s)w(s') \rangle$ is given by the delta function $\sigma_\alpha^2 \delta(s-s')$, where $\sigma_\alpha^2$ is the variance of the random input. The integral over $s'$ can be solved by using the translation property of the delta function:
\begin{equation}
\int_{-\infty}^{\infty} \delta(s - s') G (t'- s')  ds' = G (t' - s). \label{eq:delta function |, appendix}
\end{equation}

Using this formula and introducing the new integration variable $s^*$ equal to $t' -s$, the covariance function becomes
\begin{equation}
k_\alpha(t,t') = \sigma_\alpha^2 \int_{-\infty}^{\infty} G (t - t' + s^*)  G (s^*) ds^* . \label{eq:deriv. prior cov ||, methods}
\end{equation}
Since the covariance function depends on $t$ and  $t'$ only though their difference $\tau = t - t'$, we denote it as $k_\varrho (\tau)$.

\section*{Appendix II: Spherical harmonics and spherical Fourier transform } \label{sec: appendix IIs}

Spherical harmonics are the generalization of sine and cosine on the surface of a sphere. They are parametrized by the integers $l$ and $m$, of which $l$ is a positive integer and $m \in \{-l,\ldots,l\}$. These two parameters determine, respectively, the angular frequency and the spatial orientation. Spherical harmonics are defined by the following formula:
\begin{equation}
\mathcal{H}_l^m(\boldsymbol{x})= \mathcal{H}_l^m(\alpha, \theta) = \sqrt{\frac{(2l+1)(l - |m|)!}{4 \pi (l + |m|)!}} P_l^{|m|} \cos \alpha
\begin{cases}
1, & \text{for} \ m = 0\\
\sqrt{2} \cos m \theta & \text{for} \ m > 0 \\
\sqrt{2} \sin |m| \theta & \text{for} \ m < 0
\end{cases}
\, ,
\label{eq:Spherical harmonics, methods}
\end{equation}
where $P_l^{|m|}$ is a Legendre polynomial~\cite{groemer1996geometric}. 

Spherical harmonics form a set of orthonormal basis functions and, consequently, we can use them to define a spherical Fourier analysis \cite{mohlenkamp1999fast}. Specifically, the spatiotemporal  process $\alpha (x,t)$ can be expressed as a linear combination of spherical harmonics
\begin{equation}
\alpha (x,t) = \sum_{l,m} \tilde{\alpha} (l,m;t) \mathcal{H}_l^m(\boldsymbol{x}), \label{eq:inverse Fourier, methods}
\end{equation}
where $\tilde{\alpha} (l,m;t)$ is the $l,m$-th spherical Fourier coefficient as a function of time, defined as
\begin{equation}
\tilde{\alpha} (l,m; t) = \int_C \alpha (l,m;t) \mathcal{H}_l^m(\boldsymbol{x})d\boldsymbol{x}. \label{eq:direct Fourier, methods}
\end{equation}
Eqs.~(\ref{eq:inverse Fourier, methods}) and (\ref{eq:direct Fourier, methods}) are the equivalent of respectively inverse and direct Fourier transform for functions defined on the surface of a sphere.

\section*{Appendix III: Properties of the Kronecker product and GP regression with separable covariance matrices} \label{sec: appendix III}

In order to derive the posterior expectations of the spatiotemporal GP regression, it is useful to introduce some of the properties of the Kronecker product between matrices.
The Kronecker product between two $N \times N$ matrices is defined by the block form:
\begin{equation}
A \otimes B = 
\begin{bmatrix} a_{11} B & \cdots & a_{1N} B \\ \vdots & \ddots & \vdots \\ a_{N1} B & \cdots & a_{NN} B \end{bmatrix}
\, .
\label{eq:Kronecker I, app I}
\end{equation}
The following formula relates the regular matrix product with the Kronecker product:
\begin{equation}
(A \otimes B) (C \otimes D) = (AC) \otimes (BD) \label{eq:Kronecker II, app I}
\end{equation}
The inverse and transpose of a Kronecker product are respectively
\begin{equation}
(A \otimes B)^{-1} = A^{-1} \otimes B^{-1} \, .
\label{eq:Kronecker III, app I}
\end{equation}
and
\begin{equation}
(A \otimes B)^{T} = A^{T} \otimes B^{T} \label{eq:Kronecker IV, app I}\,.
\end{equation}
The following formula relates the Kronecker product to the vectorization of a matrix:
\begin{equation}
(A \otimes B) \text{vec} (C) = \text{vec} (B^T C A). \label{eq:Kronecker V, app I}
\end{equation}

Using these formulas, we can now derive the posterior expectation~(\ref{eq:posterior mean spatiotemporal, methods}) of the spatiotemporal GP regression.
Combining the spatiotemporal prior~(\ref{eq:matrix normal prior, methods}) and the observation model~(\ref{eq:matrix normal likelihood, methods}) using Bayes' theorem, we obtain the posterior
\begin{equation}
p(\text{vec} (\tilde{R}) | \text{vec} (Y)) \propto N( \text{vec} (Y)| (\Lambda \otimes I) \text{vec} (\tilde{R}), \Sigma \otimes K_\xi) N(\text{vec}(\tilde{R})|0, D \otimes K_{\rho} ) \label{eq:Derivation I, app III}
\end{equation}
This is the product of two multivariate Gaussian densities and it is therefore a multivariate Gaussian itself. Its  expectation is given by
\begin{equation}
\text{vec} (M_{\tilde{\rho}|y}) = (K_{\rho} \otimes D) (I \otimes \Lambda)^T \Bigg( (I \otimes \Lambda) (K_{\rho} \otimes D) (I \otimes \Lambda)^T + (K_\xi \otimes \Sigma) \Bigg)^{-1} \text{vec} (Y). \label{eq:Derivation II, app III}
\end{equation}
Using~(\ref{eq:Kronecker II, app I}) and (\ref{eq:Kronecker IV, app I}), the expression simplifies to:
\begin{equation}
\text{vec} (M_{\tilde{\rho}|y}) = \big(  K_{\rho} \otimes (D \Lambda^T) \big) \Bigg(  K_{\rho} \otimes (\Lambda D \Lambda^T) + (K_\xi \otimes \Sigma) \Bigg)^{-1} \text{vec} (Y). \label{eq:Derivation III, app III}
\end{equation}
This formula involves the inversion of a matrix that is the sum of two Kronecker product components. Inverting this matrix would be computationally impractical. We simplify the problem by imposing $\Sigma = \Lambda D \Lambda^T$. In this case, Eq.~(\ref{eq:Kronecker III, app I}) allows to invert the spatial and temporal covariance matrices separately:
\begin{equation}
\text{vec} (M_{\tilde{\rho}|y}) = \big(  K_{\rho} \otimes (D \Lambda^T) \big) \Bigg( \big(  K_{\rho} + K_\xi \big)^{-1}  \otimes \big(\Lambda D \Lambda^T \big)^{-1} \Bigg) \text{vec} (Y). \label{eq:Derivation IV, app III}
\end{equation}
In most realistic cases, the MEG observation model $\Lambda$ will not be full rank, therefore we introduced a Tikhonov regularization parameter $\lambda$. 
\begin{equation}
\big( \Lambda D \Lambda^T \big)^{-1} ~ \rightarrow ~ \big( \Lambda D \Lambda^T  + \lambda I\big)^{-1}
\label{regularization, app III}
\end{equation}
Using Eq.~(\ref{eq:Kronecker V, app I}), we finally arrive at Eq.~(\ref{eq:posterior mean spatiotemporal, methods}):
$$
M_{\tilde{\rho}|Y} = D \Lambda^T ( \Lambda D \Lambda^{T} + \lambda I)^{-1} Y (K_{\rho} + K_\xi)^{-1} K_{\rho}\,. 
$$

\section*{Appendix IV: Modeling vector-valued sources using block matrices} \label{sec: appendix IV} 
The source reconstruction formulae~(\ref{eq:spatial fourier posterior mean, methods}) and~(\ref{eq:spatial posterior mean, methods}) are expressed for fixed dipole directions $\vec{v}(\boldsymbol{x})$. The solution for the general case, in which the dipole direction is estimated from the data, is obtained by introducing an independent set of spherical harmonics for each of the orthogonal spatial directions $\vec{v}_1$, $\vec{v}_2$, and $\vec{v}_3$. In this Appendix, we refer to the (spherical harmonics domain) forward model matrix relative to the $k$-th direction as $\Lambda_k$. Using this notation, we can define the total forward model matrix with the following block form:
\begin{equation}
\Lambda_{tot} = 
\begin{bmatrix}
    \Lambda_1  \\
    \Lambda_2  \\
    \Lambda_3
\end{bmatrix} \label{eq: total forwardf model}
.
\end{equation}
Using an analogous notation, the total spherical harmonics covariance matrix can be written in the following block diagonal form:

\begin{equation}
D_{tot} = 
\begin{bmatrix}
    D_1&&0&&0  \\
    0&&D_2&&0  \\
    0&&0&&D_3
\end{bmatrix} \label{eq: total harmonic covariance function}
.
\end{equation}
Hence, the general source reconstruction formula in the spherical harmonics domain is obtained from Eq.~(\ref{eq:spatial fourier posterior mean, methods}) by replacing $D$ and $\Lambda$ with $D_{tot}$ and $\Lambda_{tot}$ respectively. This solution can be mapped back to the spatial domain using the total spherical harmonics matrix
\begin{equation}
H_{tot} = 
\begin{bmatrix}
    H&&0&&0  \\
    0&&H&&0  \\
    0&&0&&H
\end{bmatrix} 
\, ,
\label{eq: total spherical harmonics matrix}
\end{equation}
where $H$ is defined as in Eq.~(\ref{eq:harmonic matrix, methods}).

\nolinenumbers

\bibliography{GPbiblio4}

\end{document}